\documentclass[numberedappendix]{emulateapj}
\usepackage{apjfonts}
\DeclareMathSymbol{v}{\mathord}{cmletters}{"76}
\usepackage{amsmath}
\usepackage{amssymb}
\usepackage{hyperref}

\newcommand{\refapp}[1]{Appendix~\ref{#1}}
\newcommand{\useiop}{0}
\newcommand{\myeqref}[1]{\eqref{#1}}

\usepackage{ifthen}

\defcitealias{tchekhovskoy_ff_jets_2008}{TMN08}

\newcommand{\mybeta}{v}
\newcommand{\fpt}{{\rm fp}}

\newcommand{\thetamax}{\theta_{\rm max}}

\newcommand{\const}{{\rm const.}}
\newcommand{\gammamax}{\mu}

\newcommand{\myvec}{\vec}
\newcommand{\mytext}[1]{{\rm #1}}

\newcommand{\trm}{\infty}

\newcommand{\simplefrac}[2]{{#1}/{#2}}

\newcommand{\abs}[1]{\ensuremath{\left|#1\right|}}
\newcommand{\Abs}[1]{\ensuremath{\left|#1\right|}}

\newcommand{\ie}{i.e.\hbox{}}
\newcommand{\eg}{e.g.\hbox{}}

\newcommand{\wrt}{w.r.t.\mbox{}}
\newcommand{\cf}{c.f.\hbox{}}

\newcommand{\Swift}{\emph{Swift}}

\shortauthors{A. Tchekhovskoy, J.~C. McKinney, \& R. Narayan}
\shorttitle{Efficiency of Magnetic to Kinetic Energy Conversion in
a Monopole Magnetosphere}

\begin{document}
\label{firstpage}

\title{Efficiency of Magnetic to Kinetic Energy Conversion in a
Monopole Magnetosphere}

\author{Alexander Tchekhovskoy,$^1$
    Jonathan C. McKinney,$^2$
    Ramesh Narayan$^3$}
\altaffiltext{1}{Harvard-Smithsonian Center for Astrophysics, 60 Garden Street, MS 10,
    Cambridge, MA 02138, USA; atchekho@cfa.harvard.edu}
\altaffiltext{2}{Kavli Institute for Particle Astrophysics and Cosmology, Stanford University, P.O. Box 20450, MS 29,
Stanford, CA 94309, USA; Chandra Fellow; jmckinne@stanford.edu}
\altaffiltext{3}{Institute for  Theory and Computation, Harvard-Smithsonian Center for Astrophysics,
  60 Garden Street, MS 51, Cambridge, MA 02138, USA; rnarayan@cfa.harvard.edu}

\begin{abstract}

Unconfined relativistic outflows from rotating, magnetized compact
objects are often well-modeled by assuming the field geometry is
approximately a split-monopole at large radii. Earlier work has
indicated that such an unconfined flow has an inefficient conversion
of magnetic energy to kinetic energy. This has led to the conclusion
that ideal magnetohydrodynamical (MHD) processes fail to explain
observations of, e.g., the Crab pulsar wind at large radii where
energy conversion appears efficient. In addition, as a model for
astrophysical jets, the monopole field geometry has been abandoned in
favor of externally confined jets since the latter appeared to be
generically more efficient jet accelerators. We perform time-dependent
axisymmetric relativistic MHD simulations in order to find steady
state solutions for a wind from a compact object endowed with a
monopole field geometry. Our simulations follow the outflow for $10$
orders of magnitude in distance from the compact object, which is
large enough to study both the initial ``acceleration zone'' of the
magnetized wind as well as the asymptotic ``coasting zone.'' We obtain
the surprising result that acceleration is actually {\it efficient} in
the polar region, which develops a jet despite not being confined by
an external medium. Our models contain jets that have sufficient
energy to account for moderately energetic long and short gamma-ray
burst (GRB) events ($\sim 10^{51}$--$10^{52}$ erg),
collimate into narrow opening angles
(opening half-angle $\theta_j \approx 0.03$ rad), become
matter-dominated at large radii (electromagnetic energy flux per unit
matter energy flux $\sigma<1$), and move at ultrarelativistic Lorentz
factors ($\gamma_j \sim 200$ for our fiducial model). The simulated
jets have $\gamma_j\theta_j \sim 5$--$15$, so they are in principle
capable of generating ``achromatic jet breaks'' in GRB afterglow light
curves. By defining a ``causality surface'' beyond which the jet
cannot communicate with a generalized ``magnetic nozzle'' near the
axis of rotation, we obtain approximate analytical solutions for the
Lorentz factor that fit the numerical solutions well. This allows us
to extend our results to monopole wind models with arbitrary
magnetization. Overall, our results demonstrate that the production of
ultrarelativistic jets is a more robust process than previously
thought.

\end{abstract}

\keywords{
relativity ---
MHD ---
gamma rays: bursts ---
X-rays: bursts ---
galaxies: jets ---
accretion, accretion disks ---
black hole physics ---
methods: numerical, analytical
}

\section{Introduction}
\label{sec_intro}

Gamma-ray bursts (GRBs), active galactic nuclei (AGN), x-ray binaries,
and pulsar wind nebulae (PWNe) are among the most powerful systems
in the Universe.  Their power originates from a central engine that
contains a rotating, magnetized compact object such as a neutron
star or black hole \citep{gol69,bz77} or from a surrounding
accretion disk \citep{sha73,nt73,lov76}.  These systems obtain their angular momentum and
strong magnetic field from their environment either by advection
during their formation or through accretion which is known to amplify
any weak field by magnetorotational turbulence \citep{bh98}.
The region around the compact object is often expected to contain a highly-magnetized dipolar
magnetosphere that either threads the neutron star \citep{gol69}
or develops via accretion around the black hole \citep{bz77,nia03,mck05}.
For axisymmetric rapidly rotating systems, the
dipolar magnetosphere can be well-modeled by an approximate split-monopole field geometry
at large radii once the magnetohydrodynamically-driven (MHD-driven) outflow
has passed the so-called light cylinder (i.e. Alfv\'en surface)
and reaches a point where the flow is unconfined ~\citep{ckf99,mck06jf,mck06pulff}.
For example, astrophysical jets are typically confined by some external
medium such as a disk, disk wind, or envelope of matter.
If the jet remains highly-magnetized far from such confining media
and passes far beyond the light cylinder, then the magnetic field geometry will become
approximately monopolar (e.g. \citealt{mck06jf,tchekhovskoy_ff_jets_2008}).

The Crab Pulsar is the quintessential astrophysical object for which the
unconfined split-monopole field geometry remains a key model element.
One of the most contentious issues is how to reconcile Crab PWN observations
with MHD and pair-creation theories.
Calculations of pair formation fronts
both near the surface of the neutron star in polar gaps~\citep{sturrock71,rs75,saf78,daugherty1982pulsarcascades,hibschman2001a, hibschman2001b}
and farther from the neutron star in slot gaps~\citep{arons83} and outer gaps~\citep{crs76,chr86b,ry95,romani96,crz00}
suggest that the ratio of electromagnetic energy flux to matter energy flux in the inner pulsar wind is
\mbox{$\sigma_0\sim10^3$--$10^5$} and the Lorentz factor is
\mbox{$\gamma_0\sim10$--$1000$}. This wind is believed to terminate in a standing
reverse shock at a distance of $\sim 0.1$\,pc, i.e., at
\hbox{$\sim 3\times 10^{11}$} neutron star radii. Observations of the
shocked gas, coupled with modeling, indicate that the pre-shock plasma
has a weak magnetization, $\sigma_\trm\lesssim0.01$, which is $5$--$7$
orders of magnitude smaller than the initial magnetization
\citep{rees_crab_1974, kennel_confinement_crab_1984,
kennel_mhd_crab_1984, emmering_shocked_1987}.

How does the high-$\sigma$ wind flowing out of the star convert
essentially all of its Poynting energy flux into kinetic energy flux?
This remains an enigma, despite three and a half decades of study, and
has been coined the ``$\sigma$ problem.''  For the case of a neutron
star endowed with a split-monopole poloidal magnetic field --- a particularly
simple geometry --- it can be shown analytically that
ideal MHD processes can transfer at most $0.1$\% of the
Poynting energy flux from the Crab Pulsar to the
matter~\citep*{bes98}.  That is, the wind should remain highly magnetized out to
the distance of the termination shock.  In this model, the magnetization near
the termination shock is expected to be \hbox{$\sigma_\trm \sim
(\sigma_0\gamma_0) ^{2/3}\sim 10^4 \gg 1$}, and the Lorentz factor is expected to
be \hbox{$\gamma_\trm \sim (\sigma_0\gamma_0)^{1/3} \sim100$}
\citep{mic69,goldreich_julian_stellar_winds_1970,
camenzind_coldmhd_force_free_1986, bes98}.  This estimate of the
magnetization disagrees with the observationally inferred value of
$\sigma_\infty\lesssim0.01$, and the estimate of the Lorentz
factor is far smaller than the $\gamma_\infty\sim10^6$ inferred from
observations~\citep{kennel_mhd_crab_1984,spitkovsky_arons_2004}.

One might suspect that the above results are artificial, since they
are derived for the special case of a split-monopole geometry. However,
an approximately split-monopole is actually quite an accurate
description of the far regions of a dipolar pulsar magnetosphere, and
various studies have indicated that the low efficiency of the
split-monopole magnetosphere carries over to the dipolar
problem~\citep{ckf99,uzd_pulsar_2003, gruzinov_pulsar_2005,
mck06pulff, spitkovsky_pulsar_2006,komi_ns_06,buc06}.  This is the
reason for continued interest in the split-monopole problem.
For compactness, we hereafter refer to the case where a star
is endowed with the split-monopole magnetic field
geometry as simply the monopole magnetic field geometry case even
though the global solution away from the star is not exactly monopolar.

Various studies have explored the conditions needed for strong
acceleration of a relativistic magnetized wind and efficient conversion
of magnetic energy to kinetic energy.  \citet{camenzind_finite_element_1987},
\citet{camenz_1989}, \citet{li92}, \citet{begelman_asymptotic_1994},
and \citet{chiueh_crab_1998} showed that, for efficient energy
conversion to occur, magnetic field lines should expand away from one
another and away from the equatorial plane.  This field geometry
was identified as a ``magnetic nozzle'' because the expansion
of field lines away from the equatorial plane is geometrically
similar to the expanding outer edge of nozzles (e.g. de Laval nozzle) intended to launch a supersonic flow.
By this argument, the monopole geometry is particularly inefficient since field lines are perfectly
radial~(\citealt{bes98,bog01}, \citealt*{komi_ns_06,buc06,
barkov_ultrarel_proceeding_2008,komi_grb_jets_2008}).  Field geometries other than
monopolar/dipolar do manage to convert Poynting flux to kinetic energy
flux more efficiently, reaching
$\sigma\lesssim1$~(\citealt*{li92,begelman_asymptotic_1994, vla03a,
vla03b,vla04, bes06}, \citealt{barkov_ultrarel_proceeding_2008,komi_grb_jets_2008}).
However, all the cases considered so far that show efficient acceleration
to large Lorentz factors ($\gamma\gtrsim100$),
have involved outflows that were
restricted to flow inside collimating walls with prescribed shapes or
confining pressure profiles that induce collimation.
Some prior ideal MHD work claiming to solve the $\sigma$-problem
prescribed the field line shape, \ie, did not have a self-consistent (global force-balanced)
solution~\citep{takahashi_sigma_problem_1998,con02,fo04}.
It remains unknown whether these ``jet'' models continue to exhibit efficient
acceleration if the walls are either removed or given a different
shape or if the field line shape is self-consistently computed.

By considering small perturbations to the
monopole field geometry, \citet{bes98} derived self-consistent solutions
of highly magnetized monopole outflows near the midplane and found
inefficient acceleration.  \citet{lyub_eichler_2001} have extended their
analysis to the polar regions of such outflows and showed that
highly-magnetized ($\sigma\gg1$) collimated relativistic
jets can form there, however, they did not explore
whether these jets can become matter-dominated ($\sigma<1$) at relevant distances.
\citet{bog01,komi_ns_06,buc06,komi_grb_jets_2008}
have numerically simulated unconfined magnetized outflows
and confirmed their low efficiency at converting magnetic to kinetic energy,
\ie\ the outflows remained highly magnetized $\sigma\gg1$ out to the
simulated distances.
\citet{tomimatsu_relativistic_cone_2003} found solutions to cold ideal MHD jets
that were limited to lie inside very narrow boundaries with $\theta_j\ll 1/\sigma_0$
and in which the poloidal curvature force was neglected \citep{vla04}.
Recently
\citet*{zbb08} studied  conversion of internal energy to kinetic energy
in hot ideal MHD jets, assuming a purely toroidal magnetic field
and also assuming self-similarity that does not allow for
efficient conversion of magnetic to kinetic energy. We note that studies
of highly magnetized flows in the force-free approximation
(which neglects matter inertia and kinetic energy, \S\S\ref{subsec_M90} and \ref{sec_fms})
have given much insight into how jets/winds are launched and
into their structure~(\citealt{camenzind_finite_element_1987,
appl_camenzind_asymptotic_1993,con95,fendt97,lovelace_poynting_jets_2003,
lovelace_jet_pulsar_2006,um06,um07,mck07b,nar07,tchekhovskoy_ff_jets_2008}).

In addition to the above studies, various models have been proposed that
involve dissipative processes, \eg\ reconnection
\citep{lyutikov_uzdensky_relativistic_reconnection_2003,
zenitani_reconnection_2008,malyshkin_hall_reconnection_2008,uzd_hallmhd_2009},
as possible resolutions to the
$\sigma$-problem. In a striped wind model
\citep{michel_striped_wind_1971,coroniti_striped_wind_1990,
michel_pulsar_wind_1994} reconnection in a warped equatorial current
sheet converts magnetic energy into the kinetic energy of the plasma.
However, it remains uncertain whether such a reconnection process is
fast enough to accelerate the plasma as
required~\citep{lyubarsky_striped_wind_2001, kirk_striped_wind_2003,lyu05}.
More recently, \citet{petri_lyub_2007} have shown
that magnetic reconnection of the warped equatorial current sheet may occur right
at the wind termination shock, leading
to a decrease in the inferred pre-shock wind magnetization.
\citet{begelman_pinch_instability_1998} suggested that toroidal field
instabilities may lead to dissipation of the toroidal field,
thus circumventing the arguments by~\citet{rees_crab_1974} and
\citet{kennel_confinement_crab_1984} that inferred a low value of
$\sigma$ in the termination shock by assuming the shock
contains an ordered and purely toroidal field.  New analysis
is required to check consistency between observations and MHD models
involving disordered toroidal fields due to MHD instabilities.
However, even if linear MHD instabilities are present, as argued by
\citet{begelman_pinch_instability_1998}, their effectiveness remains unknown since they may
evolve to a saturated non-linear state that has negligible dissipation,
e.g.\ as demonstrated recently for outflows from black holes by \citet{mb08}.
We note also that \citet*{nlt09} showed for a
simple jet configuration that the linear instability growth rate
is much lower than one might expect from standard instability
criteria.

In this paper, we present a detailed study of the relativistic
magnetized monopole wind using both numerical and analytical \emph{ideal}
relativistic MHD methods.
The simulations we report here involve a much larger dynamic range
than any previous published work; they extend in radius from $r=1$,
the surface of the neutron star, to $r=10^{10}$.  The wide range of
radius allows us to study the solution far into the asymptotic region
of the wind where acceleration has practically ceased.  Also, we
consider a number of different prescriptions for the mass-loading of
field lines at the stellar surface.

The goal of this study is two-fold.  First, we wish to focus on field
lines in the equatorial region of the outflow to study the classic
$\sigma$ problem.  In particular, we wish to compare numerical results
with previously published analytical results for the asymptotic
Lorentz factor and magnetization parameter (e.g., \citealt{bes98}).
Second, we wish to study the behavior of field lines near the rotation
axis.  Even though the monopole problem is highly idealized,
nevertheless, we believe the polar field lines in this model may be
viewed as analogs of relativistic jets and indeed may even be directly
relevant to relativistic jets that become unconfined at large radii.
Our goal is to understand if there are any limitations on acceleration along polar
field lines.  In other words, is there a $\sigma$ problem for jets?

In \S\ref{sec_setup}, we describe the problem setup and the numerical
method we use to carry out the simulations.  In \S\ref{sec_results},
we present results for two simulated models: M90 and M10.  In
\S\ref{sec_analyt}, we study the shapes of field lines and explore the
connection between field line shape and acceleration.  We show that
there is a large difference between equatorial and polar field lines.
In \S\ref{sec_acceff}, we investigate what role if any is played by
signals traveling from one region of the magnetosphere to another, and
how this affects the efficiency of acceleration.  Once again, we
discover that equatorial and polar field lines have qualitatively
different rates of acceleration and efficiency. We discuss the implications of our results
in~\S\ref{sec_discussion}, and conclude in \S\ref{sec_conclusions}.

We work throughout with Heaviside-Lorentz units, and we set the
speed of light, the radius of the central compact object and the
radial component of the surface magnetic field to unity.  We use
spherical polar coordinates, $r$, $\theta$, $\varphi$, as well as
cylindrical coordinates, $R=r\sin\theta$, $\varphi$, $z=r\cos\theta$.

\section{Problem Setup}
\label{sec_setup}

\subsection{Initial Conditions and Time Evolution}

We idealize the central neutron star as a perfectly conducting sphere
that we refer to as the ``star.'' We assume that the star has a
split-monopole magnetic field configuration, with unit field strength
at the stellar surface (this choice sets the energy scale).
Exterior to the star, we initialize the system with a low but finite
rest-mass density atmosphere, which is done because the code cannot
accurately evolve a large density contrast between the initial and
injected density or a large $\sigma$ value in the initial atmosphere.
The density of the atmosphere is chosen so that it is
dynamically unimportant (the kinetic energy of the piled-up atmosphere is
much less than the kinetic energy of the wind). This atmosphere
is easily swept away by the outflowing MHD wind and has no effect on
the final results. This was confirmed by considering otherwise
identical models but where the atmosphere density was $20$ times lower.
We find all our results are converged indicating
negligible impact by the atmosphere on the injected wind.

The initial system has no rotation, so field lines are perfectly
radial and both $B_\theta$ and $B_\varphi$ vanish.  Starting with this
initial configuration, we impose a uniform rotation on the star and
study the time evolution of the external magnetosphere.  As the star
spins up, the footpoints of magnetic field lines are forced to rotate,
and this generates a set of outgoing waves traveling at nearly
the speed of light.  A short distance behind the outgoing wavefront,
the magnetosphere settles down to a steady state.  We are interested
in the properties of this steady MHD wind.

The computational domain in our simulations is the upper hemisphere,
$0<\theta<\pi/2$, with radius extending from the surface of the star,
$r=1$, to an outer edge at $r=10^{12}$.  We note that
most calculations in the literature are limited to a very small radial range due
to the need to always resolve the time-dependent compact object (e.g. \citealt{mck07a}).
Our choice of a very large range of radius allows us to study both the initial ``acceleration
zone'' of the magnetized wind as well as the asymptotic ``coasting
zone.''

\subsection{Boundary conditions}

At the polar axis, $\theta = 0$, and at the midplane, $\theta =
\pi/2$, we use the usual antisymmetric boundary conditions. At the
outer radial boundary ($r=10^{12}$) we apply an outflow condition.  At
the stellar surface, we set the poloidal component of the $3$-velocity
of the wind to a fixed value $v_p$ directed along the poloidal
magnetic field $\myvec B_p$.  We choose $v_p=1/2$ in all the
simulations reported here.  Also, we choose the angular velocity of
the star to be $\Omega=3/4$ (i.e. $v_\varphi=3/4$ at the stellar equator).
These boundary conditions are the same
for all simulations.

We assume that the magnetic field is frozen into the star.  The
radial component of the field is continuous across the stellar
surface, so we enforce the boundary condition $B_r = 1$ at $r=1$.
Since we inject a sub-Alfv\'enic flow at the surface of the rotating star,
Alfv\'en and fast magnetosonic waves communicate from the
magnetosphere back to the surface and generate self-consistent
non-zero values of $B_\theta$ and $B_\varphi$.
The fluxes at the stellar surface are set by using
``outflow''-type boundary condition (i.e., flowing out of the
computational domain into the star) on $B_\theta$ and $B_\varphi$.

Since we fix $v_p$ and $\Omega$ at the stellar surface, the toroidal
component of the $3$-velocity $v_\varphi$ is determined by the
condition of stationarity:
\begin{equation}
v_\varphi = \Omega R + B_\varphi v_p/B_p.
\label{eq_bc_vphi}
\end{equation}
This equation follows by decomposing the wind velocity into rotation
with the field line (the first term) plus motion parallel to the field
line (the second term).

The final boundary condition at the stellar surface is the plasma
density.  This controls how much mass is loaded onto field lines at
their footpoints.  The different simulations we report in this paper
correspond to different choices for $\rho(\theta_\fpt)$, the profile
of density as a function of polar angle $\theta_\fpt$ of field line
footpoints across the stellar surface.

\subsection{Numerical Approach}
\label{subsec_num}

\begin{table*}
\caption{Simulation Parameters}
\begin{center}
\begin{tabular}{@{\,}c@{\quad}c@{\quad}c@{\quad}c@{\quad}c@{\quad}c@{\quad}c@{\quad}c@{\;\;}c@{\;\;}c@{\quad}c@{\,}c@{\,}}
\hline
Name & $\theta_\mytext{max}$[$^\circ$] & $\mu_\mytext{max}$ & $\mu_\mytext{out}$ & $r_0$ & $\tilde r$ & $c$ & $n$ & $\tilde\theta$[$^\circ$] & $\tilde x_2$ & Resolution & ~~Eff. Resolution \\
\hline
\hline
\multicolumn{12}{c}{Constant density-on-the-star models}   \\
M90 &   $90$   & $460$    &    ---   & $0$    & $5$ & $0.25$ & $4$ & $11.25$ & $1/4$   & $1536$x$384$ & $3918$x$768$      \\
1000M90 %
    &   $90$   & $1000$   &    ---   & $0$    & $100$ & $0.25$ & $3$ & $90$ & $1$   & $2048$x$384$ & $2784$x$384$      \\
\hline
\multicolumn{12}{c}{Variable density-on-the-star models}  \\
M45 &   $45$   & $460$   &   $40$   & $0$    & $100$ & $0.25$ & $3$ & $90$ & $1$ & $2048$x$384$ & $2784$x$384$      \\
M20 &   $20$   & $460$   &   $40$   & $0.55$ & $100$ & $0.05$ & $3$ & $20$ & $1/3$ & $3072$x$384$ & $3216$x$576$      \\
M10 &   $10$   & $460$   &   $40$   &  $0$   & $5$ & $0.25$ & $4$ & $10$   & $1/3$ & $3072$x$384$ & $7837$x$1152$      \\
\hline
\multicolumn{12}{c}{Variable density-on-the-star + wall at $\theta = \theta_{\rm max}$}  \\
W10 &   $10$   & $460$   &   $40$   &$0.7$  & $100$ & $0.25$ & $3$ & $90$ & $1$ & $2048$x$128$ & $2786$x$128$      \\
W5 &    $5$    & $460$   &   $40$   &$0.55$ & $100$ & $0.25$ & $3$ & $90$ & $1$ & $6144$x$128$ & $8358$x$128$      \\
\hline
\end{tabular}
\end{center}
\label{tbl_models}
\end{table*}

We solve the time-dependent axisymmetric equations of special
relativistic MHD (ignoring gravity)
at zero temperature with second order accuracy.
For this we use the
code HARM \citep{gam03,mck04, mck06jf} with recent improvements
(\citealt{mm07, tch_wham07,tchekhovskoy_ff_jets_2008}).
We use HARM's second-order MC limiter method for spatial interpolations
and a second-order Runge-Kutta time-integration.  To improve the accuracy
of the simulation, before each reconstruction step we interpolate
the ratio of the numerical to approximate analytical solution,
as described in \refapp{sec_aux_approx_sol}.
To speed up the computations, we stop evolving regions of the solution where the
wind has achieved a steady state (\cf\ \citealt{kom07,tchekhovskoy_ff_jets_2008}).  This technique
allows a large gain in speed of up to a factor $\sim10^{10}$,
proportional to the maximum radius of the simulation.  Also, since our
interest is in cold flows in which the plasma internal energy and
pressure are negligibly small, we set these quantities (and all their derivatives)
identically to zero and ignore the energy evolution equation.  For example,
the conversion of conserved to primitive quantities is performed identically
to the thermal case \citep{mm07}, but pressure and internal energy and all their
derivatives are set to zero.

HARM is a flexible code that permits the use of an arbitrary
coordinate system.  We employ a radial grid in which the resolution is
very good near the star, but the cells become much more widely spaced
at large radii.  In terms of a uniformly-spaced internal code coordinate $x_1$,
we write the radial coordinate of the grid as
\begin{equation}
r(x_1) = r_0 +
       \exp\left[x_1+c H(x_1-\tilde x_1) \times (x_1-\tilde x_1)^n\right],
\label{radialgrid}
\end{equation}
where $H(x)$ is the Heaviside step function.  The lower (upper) $x_1$
value is determined by the lower (upper) radial edge of the grid: $r=1$ ($r=10^{12}$).
The parameters $r_0$, $\tilde r = r(\tilde x_1)$, $c \ge 0$, and $n > 2$ allow us to control
how resolution varies with $r$.  For instance, at $r_0 \ll r < \tilde
r$ the grid is logarithmic, with near-uniform relative grid cell size
$\Delta r/r \approx \const$  At $r = \tilde r$ the grid smoothly
switches to hyper-logarithmic, with $\Delta r/r \propto \ln^{1-1/n}r$
for $r\gg \tilde r$.  Using a hyper-logarithmic grid at large radii is
sufficient since all non-trivial changes in quantities (\eg\, the
Lorentz factor and collimation angle) occur logarithmically slowly with $r$.

For the angular coordinate, we map $\theta$ to a uniform code
coordinate $x_2$, which goes from $0$ to $1$, according to
\begin{equation}
\theta(x_2) =  x_2 \frac{\tilde\theta}{\tilde x_2}
            + H(x_2-\tilde x_2) \left(\frac{\pi}{2} - \frac{\tilde\theta}{\tilde x_2}\right)
              \times \left(\frac{x_2-\tilde x_2}{1-\tilde x_2}\right)^7.
\end{equation}
In this grid, a fraction
$\tilde x_2$ (typically $\sim1/2$) of the cells are distributed
uniformly between $\theta=0$ and $\theta=\tilde\theta$ and the remaining
cells are distributed non-uniformly at larger angles.

In different models we utilize grids with different values of the
parameters $r_0$, $\tilde r$, $c$, $n$, $\tilde x_2$, and $\tilde\theta$.
Table~\ref{tbl_models} gives the details and estimates the
effective resolution of our models in terms of the typically-used
uniform angular grid and exponential radial grid (i.e. $r(x_1)=r_0 + \exp[x_1]$).

\section{Simulation Results}
\label{sec_results}

\subsection{Baseline Model M90}
\label{subsec_M90}

\begin{figure*}
\begin{center}
\ifthenelse{\equal{\useiop}{0}}{%
\includegraphics[width=0.47\textwidth]{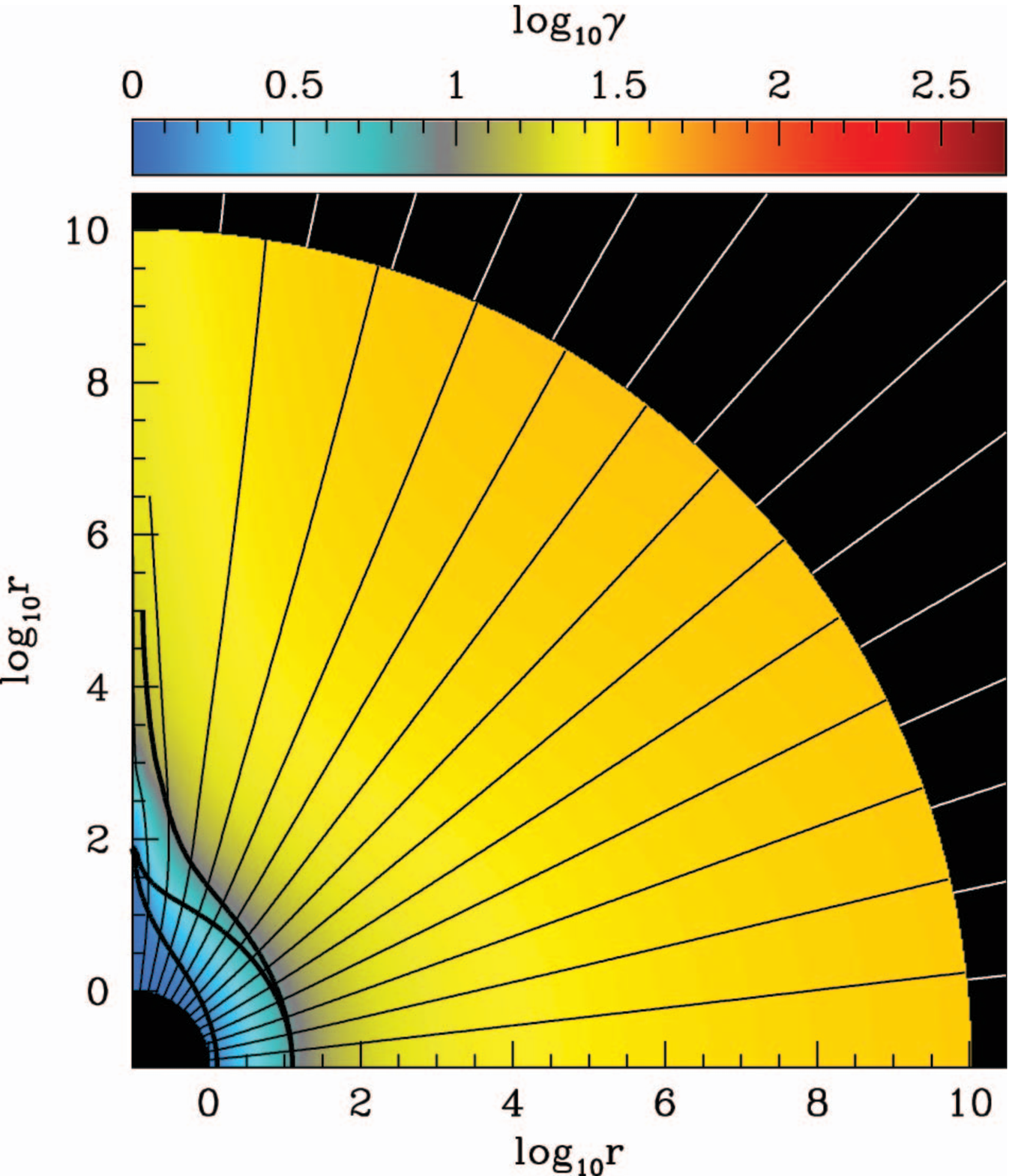}\hfill%
\includegraphics[width=0.47\textwidth]{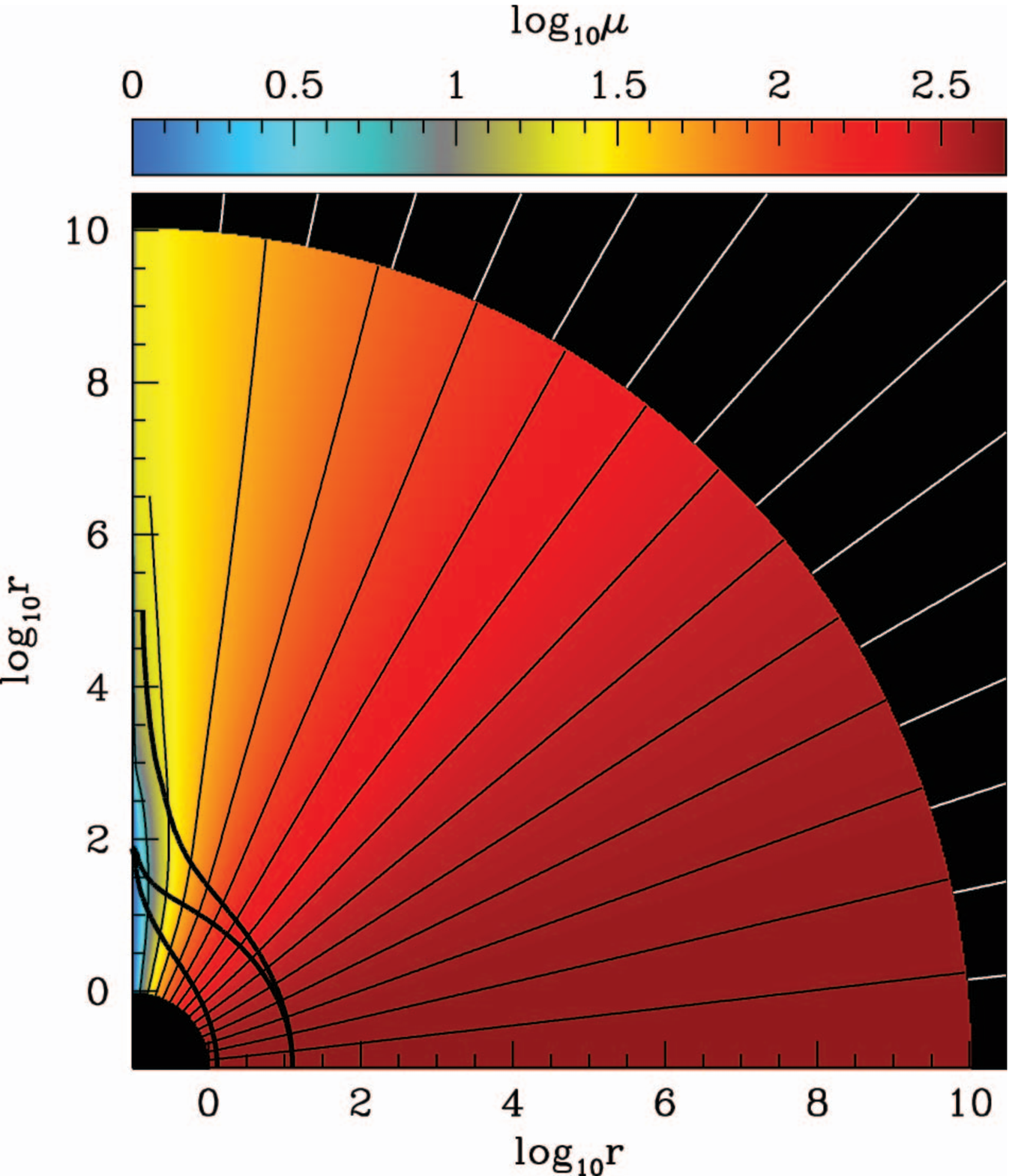}\hfill\\%
\vspace{0.5cm}
\includegraphics[width=0.47\textwidth]{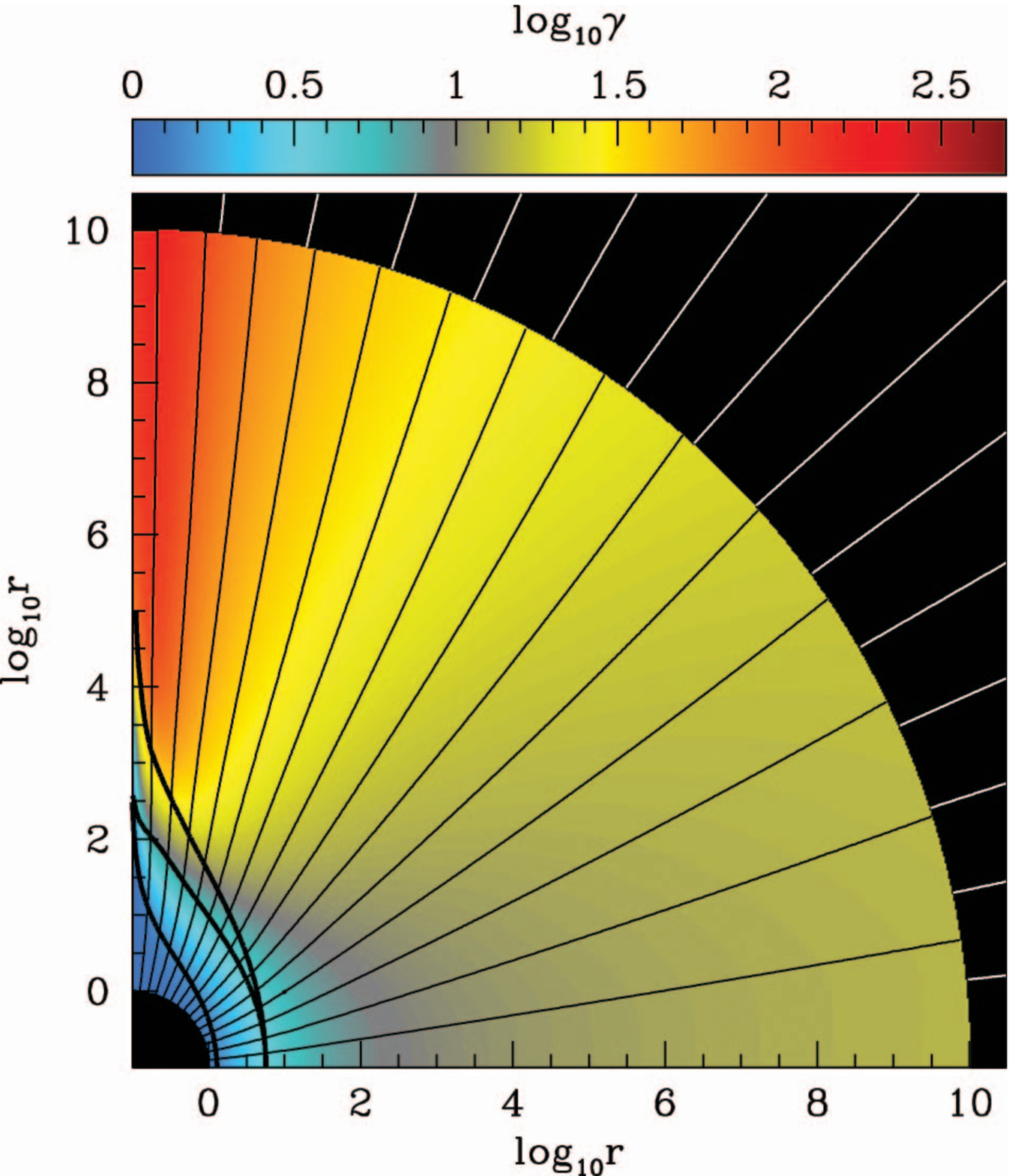}\hfill%
\includegraphics[width=0.47\textwidth]{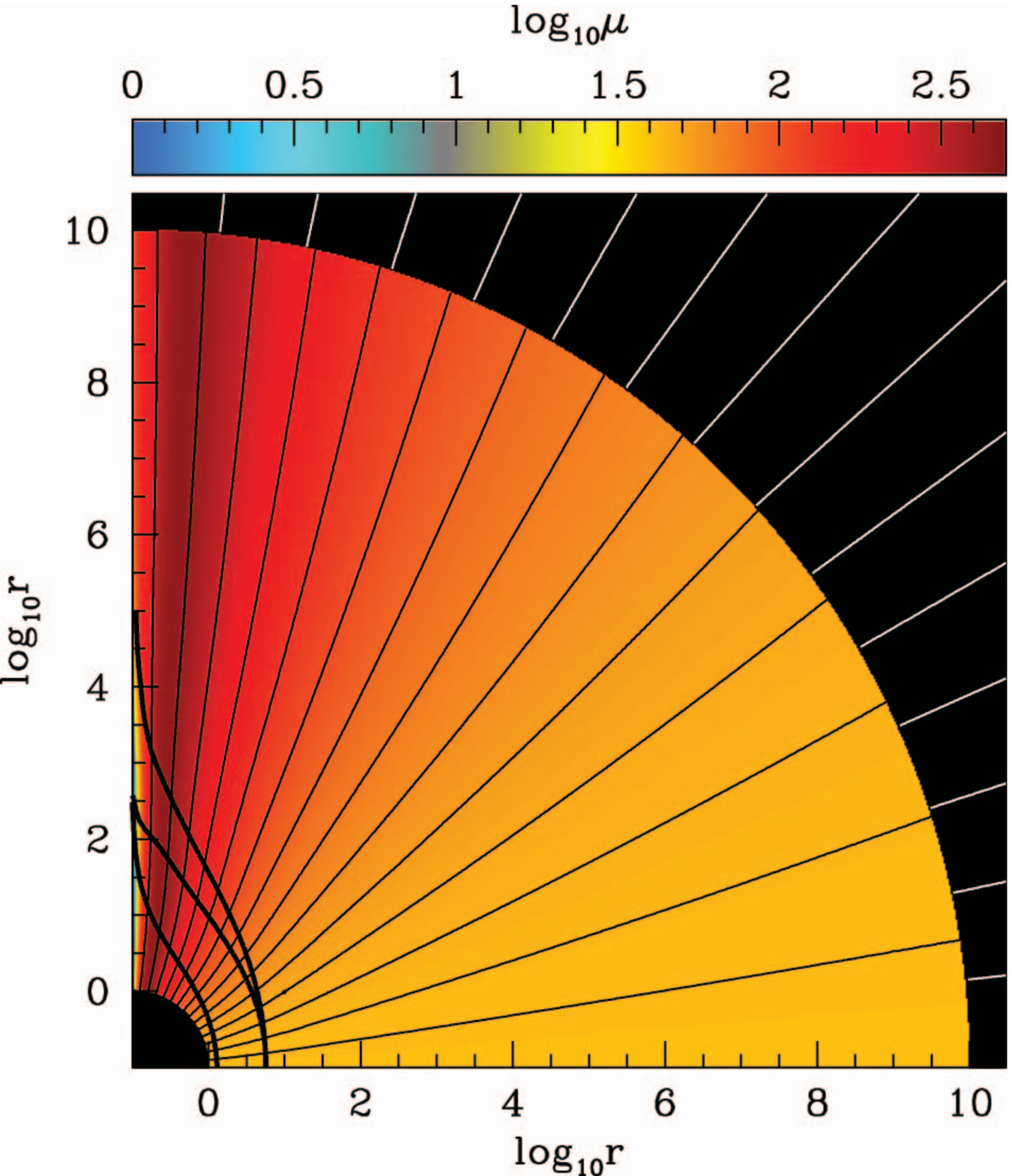}%
}{%
\includegraphics[width=0.43\textwidth]{fig1a.eps}\hfill%
\includegraphics[width=0.43\textwidth]{fig1b.eps}\hfill\\%
\vspace{0.2cm}
\includegraphics[width=0.43\textwidth]{fig1c.eps}\hfill%
\includegraphics[width=0.43\textwidth]{fig1d.eps}%
}
\end{center}
\caption{Results for models M90 (upper panels) and M10 (lower panels).
Radial dark lines show, for the steady state solution, the poloidal
structure of field lines from $\log_{10}r=0$ (the surface of the star)
to $\log_{10}r=10$.  The white radial lines on the outside show the
initial non-rotating monopolar configuration of field lines.  The
three thick solid lines correspond to the Alfv\'en surface
(innermost), the fast surface (middle), and the causality surface
(outermost, see \S\ref{sec_causality}).
The fast surface and the causality
surface touch each other at the midplane ($\theta=\pi/2$).
Top-left: color-coded Lorentz
factor $\gamma$ in model M90.  Top-right: color-coded energy flux per
unit mass flux $\mu$ in M90.  Bottom-left: $\gamma$ in M10.
Bottom-right: $\mu$ in M10. In the figure we use ``logarithmic'' spherical
polar coordinates ($r_l=1+\log_{10}r\,,\theta_l=\theta$).
The numbers along the horizontal (vertical) axis correspond to the values of
$\log_{10}r$ along that axis, \ie, $r_l \cos\theta_l - 1$
($r_l\sin\theta_l - 1$). Even though in these ``logarithmic'' polar
coordinates the flow appears to over-collimate toward the axis (\eg,
the two leftmost field lines on the upper panels),
in fact $dR/dz$ remains positive everywhere in the solution.
We note that the shape of the shown portion of the Alfv\'en surface is very close
to a cylinder $R = 1/\Omega = 4/3$.}
\label{fig_1}
\end{figure*}

For our fiducial baseline model, called M90, we set the density
$\rho(\theta_\fpt)$ at the surface of the star to a constant value
equal to $0.001914$, independent of $\theta$, such that $B_r^2/(\rho_0 c^2)\approx 522$.
The rotation of the star launches a relativistic magnetized wind that accelerates the
input mass.  The simulation is run for a time $10^{12}$ in
units of the light-crossing time of the star.  By this time, the wind reaches
a steady-state solution that is independent of initial transients
out to a radius $10^{10}$, so all our results are reported out to this radius.
The top two panels in Fig.~\ref{fig_1} show the results.

Consider first the shapes of the field lines in the poloidal plane.
We see that, over most of the solution, the field lines are only
slightly perturbed from their initial purely radial configuration.
This might be surprising since rotation causes the toroidal
component of the field to grow substantially.  In fact, over most of
the solution, $B_\varphi$ is tens of times larger than the poloidal field
$B_p\equiv\sqrt{B_r^2+B_\theta^2}$, and one might think that the hoop
stress of the strong toroidal field would cause substantial
collimation of the field lines.  This does not happen because the
relativistic wind sets up an electric field, and an associated
electric force per unit volume $\rho_e\myvec{E}$ (charge density $\rho_e$),
that almost exactly cancels the magnetic hoop stress.  This is a unique feature of
relativistic winds.  It has been explored in detail
by~\citet{nar07} and \citet[hereafter \citetalias{tchekhovskoy_ff_jets_2008}]{tchekhovskoy_ff_jets_2008} in
connection with winds in the force-free approximation \citep{mck06ffcode,mck07b}.

While it is true that the distortion of field lines in the poloidal
plane is small, nevertheless, there is some distortion.  It is
especially obvious near the rotation axis, where we see field lines
converging towards the pole and bunching up.  In fact, very close to the
rotation axis, the distortion appears to be quite large.\footnote{Note
that Fig.~\ref{fig_1} uses a logarithmic radial coordinate and thus exaggerates the
effect.  The actual change in $\theta$ of a field line is not very
severe even at very large distances from the star.}
Thus, the results indicate that the polar
regions of a rotating monopole wind are qualitatively different from
the equatorial regions of the outflow.

We next consider the acceleration of the wind.  The top left panel of
Fig.~\ref{fig_1} shows the Lorentz factor $\gamma$ as a function of
position.  We see that there is an inner acceleration zone extending
out to a radius $r\sim10^2$, where the Lorentz factor increases from
its initial small value ({}$\approx1.2$ at the stellar surface).
Beyond this we find a large coasting zone where there is very
little change in $\gamma$.  Over most of the coasting zone, the
flow reaches ultrarelativistic Lorentz factors of $\gamma\sim40$;
however, this value is much less than expected if all the available
free magnetic energy were to be used.

\begin{figure*}
\begin{center}
\includegraphics[width=0.8\textwidth]{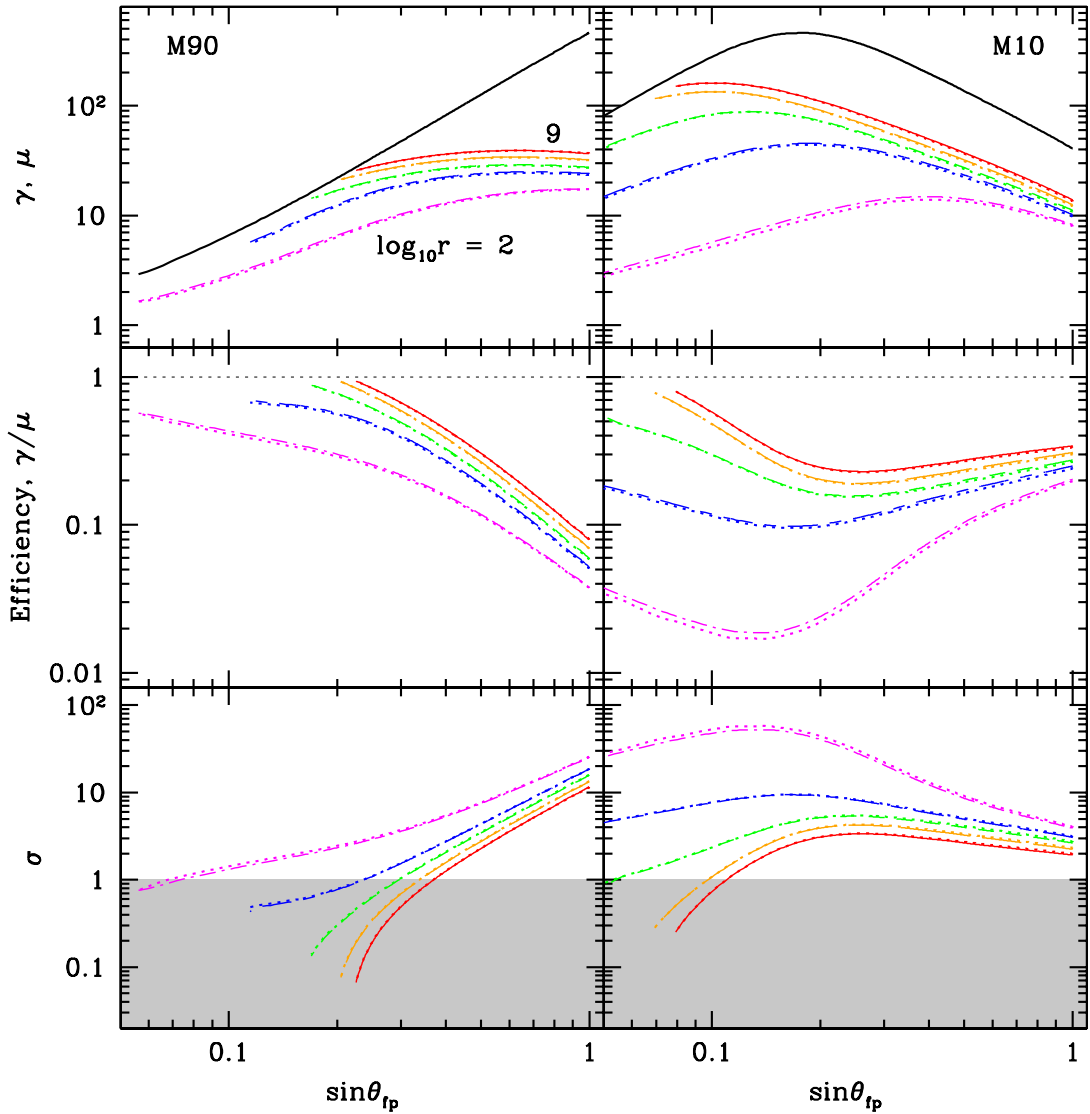}
\end{center}
\caption{The panels on the left correspond to model M90 and those on
the right to M10.  The solid black lines in the top panels show the
scaled total energy flux $\mu$
as a function of $\sin\theta_\fpt$, where $\theta_\fpt$ is the polar
angle corresponding to footpoints of field lines.  The five colored
lines show profiles of $\gamma$ at five different distances from the
star: $r=10^2$ (magenta, dash--dotted), $10^3$ (blue,
long-short--dashed), $10^4$ (green, short--dashed),
$10^6$ (orange, long--dashed), $10^9$ (red, solid).
The closer $\gamma$ gets to $\mu$ the more efficient the acceleration.
The middle panels
show the acceleration efficiency $\gamma/\mu$ and the lower panels
show the magnetization parameter $\sigma=(\mu/\gamma)-1$ at the same five distances.
Note that $\gamma/\mu$ is largest and $\sigma$ is smallest closest to
the pole ($\sin\theta_\fpt\ll1$).  The various dotted lines correspond
to the analytical model described in \S\ref{sec_analyt_aoafp}. The
shaded area in the bottom panel indicates the regions of the solution
where mass energy flux exceeds Poynting flux ($\sigma<1$, or $\gamma/\mu > 1/2$).
Note that magnetic flux gradually converges toward the polar axis as a function
of increasing radius, so the lines corresponding to larger radii
truncate at small values of $\theta_\fpt$.
}
\label{fig_2}
\end{figure*}

An axisymmetric magnetized wind has several conserved quantities along
field lines.  Two of these are the enclosed magnetic flux $\Phi$ and
the angular velocity $\Omega$.  Another is the ratio of poloidal
magnetic flux to rest-mass flux \citep{chandrasekhar_mhd_axisymm_1956,
mes61,li92,bes97}:
\begin{equation}
\eta(\Phi) = \frac{\gamma \rho v_p}{B_p} = \const \mytext{\ along\ field\ line.}
\label{eq_etadef}
\end{equation}
Yet another conserved quantity is the quantity $\mu$, which is the
ratio of the total energy flux to the rest-mass flux
\citep{chandrasekhar_mhd_axisymm_1956,mes61,
lovelace_1986,begelman_asymptotic_1994,bes97,chiueh_crab_1998}:
\begin{equation}
\mu(\Phi) = \frac{\mathcal S+\mathcal K}{\mathcal R} = \frac{E
    \abs{B_\varphi} + \gamma^2 \rho v_p}{\gamma \rho v_p} =
    \mytext{const.\ along\ field\ line}.
\label{eq_mudef}
\end{equation}
Here, ${\mathcal S}$ is the Poynting flux, ${\mathcal K}$ is the mass
energy flux (rest mass $\mathcal R$ plus kinetic energy), $B_\varphi$ is the
toroidal field,
\begin{equation}
E=\Omega R B_p
\label{eq_edef}
\end{equation}
is the poloidal electric field, and $\rho$ is the mass
density in the comoving frame of the fluid.  The denominator of
equation (\ref{eq_mudef}) is the rest mass flux.  Since we consider
highly magnetized winds, we have ${\mathcal S} \gg {\mathcal K}$ at
the surface of the star.  Moreover, at $r=1$ in the monopolar
flow \citep{mic69,mic73},
\begin{equation}
\abs{B_\varphi} \approx E \approx \Omega \sin\theta_\fpt,
\label{eq_EBsurface}
\end{equation}
where $\theta_\fpt$ represents
the value of $\theta$ at the footpoint of a field line.  Also, at the
footpoint, $v_p$ and $\rho$ are constant, and $\gamma\approx1$.  We
thus have\footnote{In this paper we label field lines by either the
enclosed magnetic flux $\Phi$ or the polar angle at the footpoint
$\theta_\fpt$.}
\begin{equation}
\mu(\theta_\fpt) \propto \sin^2\theta_\fpt,
\end{equation}
for $\mu\gg\gamma_0$ ($\gamma_0$ is the initial Lorentz factor at $r=1$),
i.e., $\mu$ is a rapidly increasing function of $\theta_\fpt$.
This can be seen in the top right panel in Fig.~\ref{fig_1}
and also in Fig.~\ref{fig_2}.

As mentioned above, the quantity $\mu$ is conserved along each field
line.  However, the two energy contributions to $\mu$, the Poynting
flux ${\mathcal S}$ and the mass energy flux ${\mathcal K}$, are not
individually conserved.  In fact, the outflowing wind converts
Poynting flux to mass energy, thereby accelerating the wind and
causing $\gamma$ to increase.

If the wind were maximally efficient at accelerating the matter, we
would expect ${\mathcal S}\to 0$ at large distance from the star. The
Lorentz factor would then achieve its maximum value
\begin{equation}
\gamma_{\rm max} = \mu.
\label{eq_gammamax}
\end{equation}
In practice, the wind falls far short of this maximum.  As Figs.~\ref{fig_1} and
\ref{fig_2} show, $\mu$ is quite large in model M90, with a value of 460 at the
equator ($\theta_\fpt=\pi/2$).  However, the Lorentz factor of the
wind, even at a radius of $10^{10}$, does not exceed $40$.  Thus, a
magnetized monopole wind is very inefficient at accelerating the gas.
For an equatorial field line in M90, the efficiency factor
$\gamma/\mu$ is only about $0.09$, as shown in Fig.~\ref{fig_2}.

Another way of describing the efficiency of conversion of energy from
electromagnetic to kinetic form is via the \emph{magnetization}
parameter $\sigma$, which is the ratio of Poynting to mass energy flux
\citep{kennel_confinement_crab_1984,li92,begelman_asymptotic_1994,vla04,kom07},
\begin{equation}
\sigma = \frac{\mathcal S}{\mathcal K}
       = \frac{E \abs{B_\varphi}}{\gamma^2 \rho v_p}.
\label{eq_sigmadef}
\end{equation}
Substituting into equation (\ref{eq_mudef}) we see that the conserved
quantity $\mu$ is related to $\sigma$ by
\begin{equation}
\mu = \gamma (\sigma+1).
\label{eq_mu_via_gamma_sigma}
\end{equation}
The smallest value possible for $\sigma$ is zero.  Therefore, the
maximum value of $\gamma$ is $\gamma_{\rm max}=\mu$ (eq.
\ref{eq_gammamax}).

The magnetization $\sigma$ is {\it not} conserved along a field line.
At the surface of the star, where $\gamma\approx1$, we have
$\sigma\approx\mu-1$.  As the magnetized wind flows out and energy is
transferred from Poynting to matter energy, $\gamma$ increases and
$\sigma$ decreases.  An efficient wind would be one in which $\sigma$
asymptotes to a value $\leq 1$, so that the outflowing material is
able to convert at least half of its energy flux into matter energy.
The numerical solutions shown in Figs.~\ref{fig_1} and \ref{fig_2}
fail to satisfy this criterion by a large factor in the equatorial
regions.  This implies there is no ideal, axisymmetric MHD solution
to the $\sigma$-problem \citep{rees_crab_1974,kennel_confinement_crab_1984,kennel_mhd_crab_1984}.

There is, however, one promising feature in the results: the polar
regions of the wind are efficient, with $\gamma/\mu\to1$ and
$\sigma\ll1$ at large distance from the star (Fig.~\ref{fig_2}).  This
interesting feature of the monopole problem has not been emphasized in
the literature.  Most previous analyses and discussions have focused
on equatorial field lines where the efficiency is, indeed,
too low to solve the $\sigma$ problem.

Unfortunately, the actual Lorentz factor along polar field lines in
M90 is only $\sim20$ since this is the value of $\mu$ for these lines.
Would we continue to have high efficiency in the polar region even
with larger values of $\mu$?  In particular, is it possible to have
acceleration with high efficiency up to Lorentz factors $\gamma >
100$, as observed for instance in gamma-ray bursts?  For this we need
to study a model with larger values of $\mu$ near the pole.  We
describe such a model in the next subsection.

\subsection{Model M10}
\label{subsec_M10}

\begin{figure}
\begin{center}
\includegraphics[width=\columnwidth]{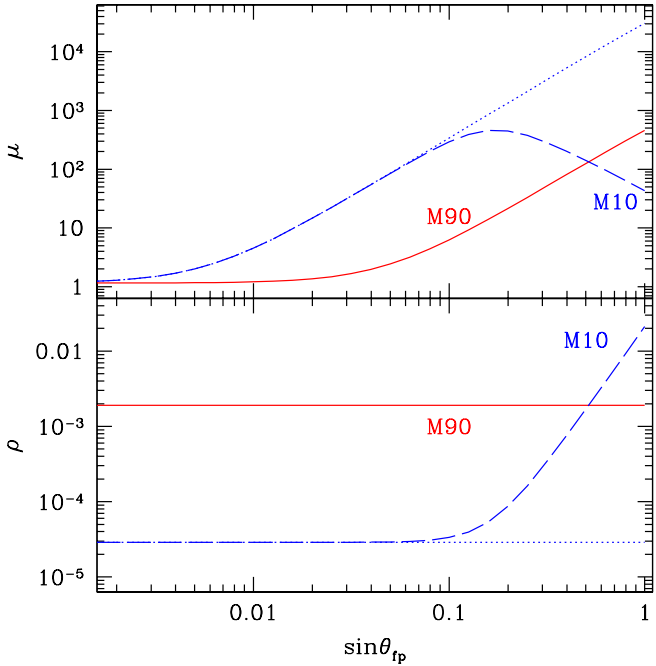}
\end{center}
\caption{The red lines in the two panels show the distribution of
$\mu$ and $\rho$ as a function of footpoint angle $\theta_\fpt$ in the
baseline model M90.  The blue dotted lines show the corresponding
profiles for a hypothetical model in which $\rho$ is decreased by a
constant factor, leading to an increase in $\mu$ by the same factor.
 Since this causes $\mu$ to become very large at the
equator, leading to numerical difficulties, we consider the model M10
(dashed blue lines) that has the same behavior of $\mu$ and $\rho$
near the axis but is less extreme near the equator.
In both models M10 and M90
the matter-dominated part of the flow $\mu \sim 1$ at the surface of the star
is resolved on the grid by at least several grid cells.
}
\label{fig_3}
\end{figure}

From equation (\ref{eq_mudef}) we see that an obvious way to increase
$\mu$ is to lower the density of the wind at the stellar surface.  For
instance, if we were to reduce $\rho$ by a factor of $\sim30$ relative
to M90, then we would have a model with $\mu\sim$\, few hundred for a
field line with $\theta_\fpt\sim10^\circ$ (see Fig.~\ref{fig_3}).  We
could then explore acceleration along this field line and determine
whether or not the outflowing wind achieves a coasting $\gamma>100$.

Since $\mu$ varies as $\sin^2\theta_\fpt$, this approach would lead to
extremely large values of $\mu$ at the equator.  As a result, the
model would require very large resolution to simulate accurately and
would be extremely expensive.  Therefore, for numerical convenience,
we consider a model in which we choose the profile $\rho(\theta_\fpt)$
such that $\mu$ is large near the pole, reaches a maximum $\mu_{\rm
max}$ at a specified footpoint angle $\theta_\fpt=\theta_{\rm max}$,
and then decreases with increasing $\theta_\fpt$ to an outer value
$\mu_{\rm out}$ at $\theta_\fpt=\pi/2$.\footnote{Decreasing $\mu$ near the equator
is actually a physically reasonable approach to model the equatorial region
if it were to contain a weakly magnetized pulsar current sheet or an accretion disk.}
To achieve this, we choose the density profile on the star to be
\begin{equation}
\rho(\theta_\fpt) = \rho_0 + \rho_1 \sin^{\alpha} \theta_\fpt,
\label{eq_bc_rho}
\end{equation}
where the constants $\rho_0$, $\rho_1$ and $\alpha$ are adjusted so
that the model has the desired values of $\mu_{\rm max}$, $\theta_{\rm
max}$ and $\mu_{\rm out}$.

\begin{figure*}
\begin{center}
\includegraphics[width=0.65\textwidth]{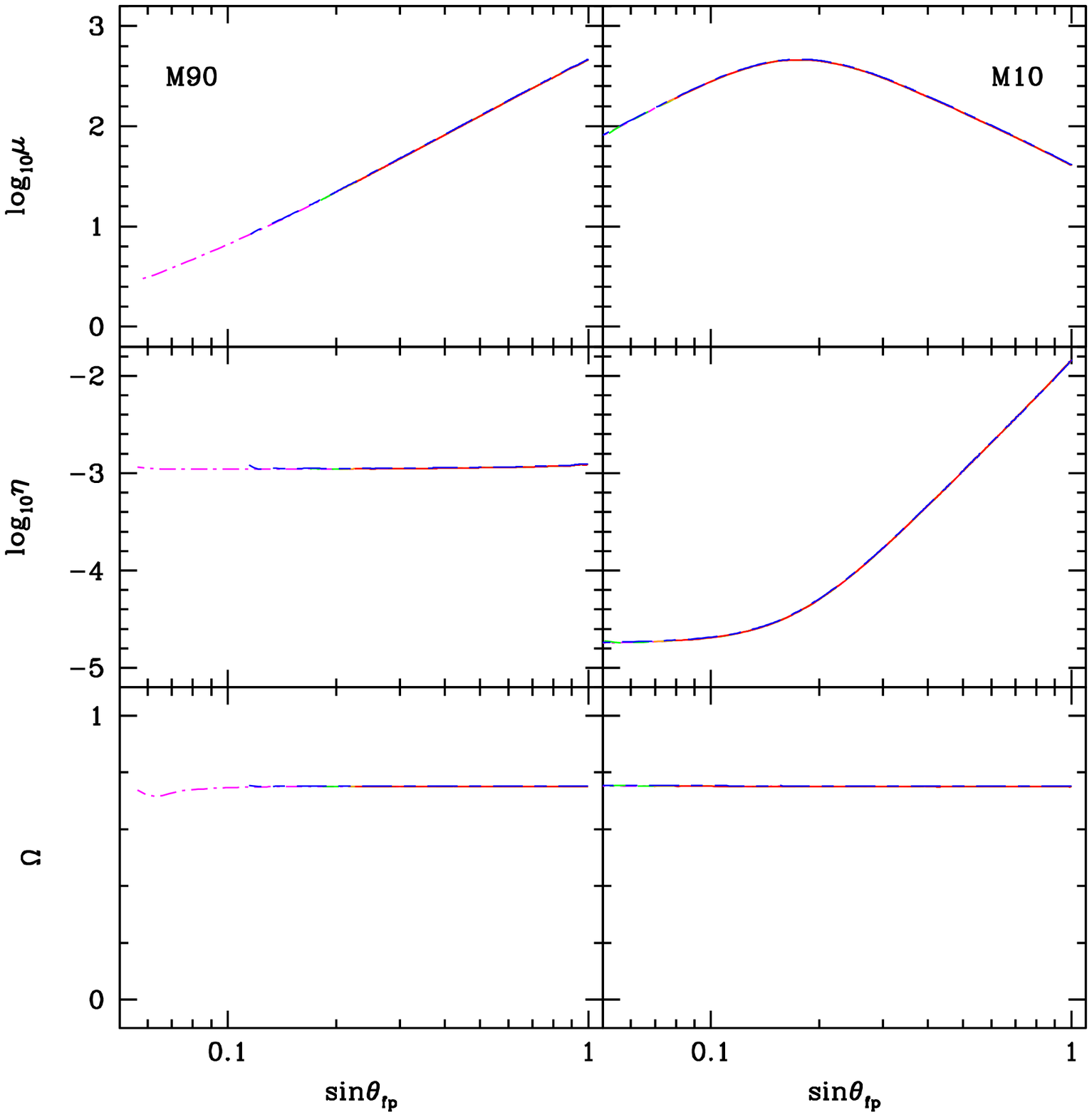}%
\end{center}
\caption{The panels on the left correspond to model M90 and those on
the right to M10.  The five colored
lines correspond to $r = 10^2$, $10^3$,
$10^4$, $10^6$, and $10^9$ (see Fig.~\ref{fig_2}
for a description of line colors and types) and show profiles of quantities
that are preserved along the field
lines: $\mu(\theta_\fpt)$ (the top panel), $\eta(\theta_\fpt)$ (the
middle panel), and $\Omega(\theta_\fpt)$ (the bottom panel).
 These lines
nearly overlap each other, as expected for a flow in steady
state. In the numerical simulations, each of the quantities is preserved
along field lines to better than $15$\%. A few grid cells
from the polar axis, within which
unresolved magnetic flux accumulates, are least accurate
(not shown); but as we have checked, this does not affect the
quality of the solution at larger angles
(see end of \S\ref{sec_comparison_with_numerical_results}).}
\label{fig_invariants}
\end{figure*}

We have simulated a series of such models, all with $\mu_{\rm
max}=460$ and $\mu_{\rm out}=40$~\footnote{We have confirmed that the
precise value we choose for $\mu_{\rm out}\gg 1$ is unimportant so
long as we are only interested in acceleration along polar field
lines.}, and with different values of $\thetamax$, viz., $45^\circ$,
$20^\circ$, $10^\circ$, which we refer to as models M45, M20, M10,
respectively. Figure~\ref{fig_3} shows the profiles of
$\rho(\theta_\fpt)$ and $\mu(\theta_\fpt)$ for the model M10.
Parameters of the various models are summarized in
Table~\ref{tbl_models}. The models are well-converged, with field-line
invariants $\Omega(\Phi)$, $\eta(\Phi)$, and $\mu(\Phi)$ conserved
along field lines to better than $15$\%, as Fig.~\ref{fig_invariants}
shows for, e.g., models M90 and M10.

The bottom two panels in Fig.~\ref{fig_1} show results corresponding
to M10.  As before, we see that the poloidal structure of the field is
largely unaffected by rotation.  There is of course some lateral shift
of field lines, the effect being larger near the pole than near the
equator, with maximum field line bunching close to the axis.

The Lorentz factor distribution again confirms the trends seen in M90.
Near the equator, the asymptotic $\gamma$ is only $\sim12$, giving an
inefficient flow with $\gamma/\mu\sim0.3$.  The efficiency increases
near the axis and becomes practically equal to unity close to the
pole; equivalently, $\sigma$ becomes much less than unity for these
field lines.  Most interestingly, Lorentz factors nearly as large as
$\sim200$ are obtained for these field lines.  In other words, there
is no $\sigma$ problem near the axis and it is possible to obtain
quite large Lorentz factors in this region of the outflow.

\subsection{Summary of Key Results}

From the results shown in Figs.~\ref{fig_1}, \ref{fig_2}, we conclude the following:
\begin{enumerate}
\item Field lines near the equator of a rotating monopole largely retain
their monopolar configuration out to large radii, whereas lines near
the pole tend to bunch up around the axis.
\item The acceleration efficiency $\gamma/\mu$ of a rotating monopole
magnetosphere is low ($\ll1$) for field lines in the equatorial region, but
quite high ($\sim1$) for field lines near the pole.
\end{enumerate}
We would like to develop an understanding of the physics behind of
these effects.  We would also like to know how the two effects are
related to each other.  This is the topic of the next two sections.

\section{Field Line Bunching}
\label{sec_analyt}

\subsection{Relation to Acceleration Efficiency}
\label{sec_analyt_aoafp}

As plasma streams along field lines, relativistic effects become
important near the so-called light cylinder, $R = R_\mytext{L} = 1 /
\Omega$, where the co-rotation velocity $\Omega R$ equals the speed of
light and $E=B_p$ (see eq.~\ref{eq_edef}).  As we show in
\refapp{sec_gamma_mhd}, far outside the light
cylinder, where $\Omega R \gg 1$ and $\gamma\gg\gamma_0$, the plasma simply drifts
perpendicular to $\myvec E$ and $\myvec B$ at the drift
velocity~\citep{bes98,beszak04,vla04},
\begin{equation}
v \approx v_\mytext{dr} = \Abs{\frac{\myvec E \times \myvec B}{B^2}} = \frac{E}{B},
\label{eq_vdrift}
\end{equation}
The corresponding Lorentz factor is
\begin{equation}
\gamma^2 \approx \gamma_\mytext{dr}^2 = \frac{B^2}{B^2-E^2}.   \label{eq_gammadrift}
\end{equation}
Near the star this formula becomes inaccurate since the plasma
moves at the initial Lorentz factor,
\begin{equation}
\gamma^2 \approx \gamma_0^2.
\end{equation}
In \refapp{sec_gamma_mhd} we show that a combination of these two formulae
does a very good job of describing the Lorentz factor
at all distances from the star:
\begin{equation}
\gamma^2 = \gamma_0^2 - 1 + \gamma_\mytext{dr}^2.
\end{equation}

In the asymptotic region of a relativistic outflow (where $\Omega R
\gg 1$, $\gamma\gg1$), we have according to \myeqref{eq_gammadrift},
\begin{equation}
\abs{B_\varphi} \approx E.
\label{eq_eb_asymptotic}
\end{equation}
Further, this relation is also true at the surface of the central
compact star (for the monopolar flow, see eq.~\ref{eq_EBsurface}).  This allows us to write
the difference between the maximum allowed Lorentz factor $\gamma_{\rm
max}=\mu$ and the local Lorentz factor $\gamma$ in the following
convenient form (the numbers in parentheses refer to the equations
used to derive this result):
\begin{equation}
\gammamax-\gamma
       \overset{\myeqref{eq_mudef},\myeqref{eq_EBsurface},\myeqref{eq_eb_asymptotic}}{\approx} 
       \frac{E^2}{\gamma\rho v_p}
       \overset{\myeqref{eq_etadef}}{=} \frac{(\Omega R B_p)^2}{\eta B_p}
       = \frac{\Omega^2(\Phi)}{\eta(\Phi)} {B_p R^2}.
\label{eq_mu_minus_gamma}
\end{equation}
By dividing this equation by itself as evaluated at the footpoint, and
approximating $\gamma - \gamma_\fpt \approx
\gamma$ and $\mu-\gamma_\fpt\approx\mu$, we obtain
\begin{equation}
\frac{\gamma}{\gammamax}
          \approx 1 - \frac{B_p R^2}{[B_p R^2]_\fpt}
          \approx 1 - \frac{a}{a_\fpt},
\label{eq_gamma_minus_mu}
\end{equation}
where we have used the subscript ``\fpt'' to denote quantities evaluated at the
field line footpoint\footnote{We note that for field geometries other than monopole,
equation~\myeqref{eq_eb_asymptotic} in general breaks down at field line
footpoints but holds at the fast
magnetosonic surface (\S\ref{sec_fms}).
Due to this reason, for field geometries other than monopolar
the subscript ``\fpt'' indicates quantities
as measured at the fast magnetosonic point.}
and have defined the quantity
\begin{equation}
a \equiv B_p R^2.
\label{eq_a}
\end{equation}

Equation (\ref{eq_gamma_minus_mu}) demonstrates that, in order to
convert an appreciable fraction of the total energy flux $\mu$ along a
field line into matter energy flux ({}$\gamma$ times the mass flux), the
quantity $a$ has to decrease appreciably from its initial value at the
footpoint.  This result is
known~\citep[\cf][]{begelman_asymptotic_1994, chiueh_crab_1998,vla04},
though we have not seen as simple a derivation as the one given above.

For a precisely monopole field, $B_p \propto 1/R^2$ along each field
line. Therefore, $a = B_pR^2$ is constant along a field line and so no
efficient acceleration is possible.  This explains why acceleration is
so difficult in the monopole problem. In order to permit acceleration,
field lines must move in a cooperative fashion transverse to one
another so as to allow $a$ to reduce with increasing distance from the
star.  We discuss how this is accomplished in the next subsection.

Meanwhile, as an aside, we describe here an improvement to the
approximate result ~\myeqref{eq_gamma_minus_mu} which gives the correct
numerical value of efficiency in a precisely monopolar field.
For such a field, the Lorentz factor at asymptotically large distances
is \citep{mic69,camenzind_coldmhd_force_free_1986}
\begin{equation}
\gamma_\trm^\mathrm{radial} = \gammamax^{1/3} \ll \mu.
\label{eq_gammamic69}
\end{equation}
This gives an extremely inefficient acceleration, $\gamma/\mu \approx
\mu^{-2/3} \ll 1$.  However, the efficiency is not zero as equation
(\ref{eq_gamma_minus_mu}) might suggest, so we need a more accurate
version of~\myeqref{eq_eb_asymptotic}.  According to
\myeqref{eq_gammadrift},
\begin{equation}
B_\varphi\approx\frac{\gamma E}{\sqrt{\gamma^2-1}}
         = \frac{\gamma\Omega R B_p}{\sqrt{\gamma^2-1}}
         \approx\Omega R B_p \left(1+\frac{1}{2\gamma^2}\right).
         \label{eq_bphi}
\end{equation}
Using this equation, assuming $B_p R^2=\const$ along a field line, and
introducing $\gamma_\infty=\lim_{R\to\infty}\gamma$, we obtain in the
limit \hbox{$R\to\infty$}
\begin{equation}
\gammamax(\Phi)
       \approx\gamma_\infty+\frac{\Omega^2(\Phi)}{\eta(\Phi)}B_p R^2\left(1+\frac{1}{2\gamma_\infty^2}\right).
\label{eq_mu_minus_gamma_acc}
\end{equation}
Now, assuming that the system settles down to a state with minimum
total energy flux $\mu(\Phi)$ (equivalent to the minimal torque
condition of \citealt{mic69}), we find the terminal Lorentz factor
$\gamma_\infty$ that minimizes the right hand side of this equation:
\begin{equation}
\gamma_\infty^\mathrm{radial} =\left[\frac{\Omega^2(\Phi)}{\eta(\Phi)}B_p R^2\right]^{1/3}
              \approx \gammamax^{1/3},
              \label{eq_gammainfty}
\end{equation}
where the approximate equality comes from equation
\myeqref{eq_mu_minus_gamma} evaluated at the footpoint. This
reproduces \myeqref{eq_gammamic69} and shows that indeed, for a
precisely radial flow, only a small (but still non-zero) fraction of
the Poynting flux is converted to the kinetic energy of the matter.
This derivation was performed for precisely monopolar field lines.  In
actuality the shape of the poloidal field lines is slightly changed
from radial.  The effect is small for equatorial field lines and the
order of magnitude estimate \myeqref{eq_gammainfty} continues to hold.
The deviations are larger for polar field lines and the Lorentz factor
obtained along these lines is very different from
(\ref{eq_gammainfty}).

\subsection{Field Lines Near the Midplane}

Let us first apply formula~\myeqref{eq_gamma_minus_mu} to a field line
near the midplane. For acceleration to be efficient, $a$ must decrease
along the field line, i.e., $B_p$ must decrease faster than $1/R^2$.
This can be accomplished by moving field lines away from the equator
towards the axis.  Consider a field line with its footpoint located at
a small angle $\theta'_\fpt =\pi/2-\theta_\fpt \ll 1$ from the
equator.  For this field line we can write \myeqref{eq_gamma_minus_mu}
as
\begin{equation}
\frac{\gamma}{\mu}\approx1-\frac{a}{a_\fpt}
                  \approx 1-\frac{\theta'_\fpt}{\theta'},
                  \label{eq_eff_near_midplane}
\end{equation}
where $\theta'$ is the polar angle of the field line at a large
distance from the star, where the Lorentz factor is $\gamma$.
Here we used the fact that
$\Phi' \approx 2\pi a\, \theta' \approx
2\pi a_\fpt\theta'_\fpt$, where $\Phi'$ is the
amount of flux enclosed between the field line and the midplane.

For a nearly monopolar configuration of the field in which
$\theta' \approx \theta'_\fpt$, obviously we will not have much
acceleration ($\gamma/\mu\ll1$). In order to obtain high acceleration
efficiency in the equatorial region, field lines must diverge from the
equator so that the $\theta'$ values of field lines increase with
distance.  For this to happen, the rest of the magnetosphere must
collectively move away from the equator towards the pole.  For reasons
that are discussed in \S\ref{sec_acceff}, this does not happen, and so acceleration
efficiency is at best modest near the equator.

\subsection{Polar Field Lines}
\label{sec_analyt_bpnonuni}

The story is quite different for field lines close to the axis
($\theta \ll 1$).  Consider the initial undistorted monopole
configuration.  At the surface of the star, $B_p$ is constant, and
$[B_pR^2]_\fpt$ is simply equal to $\Phi/\pi$, where $\Phi$ is the
flux interior to the field line.  For a pure monopole, this relation
is valid at any distance from the star, i.e., $B_pR^2=\Phi/\pi$ at all
radii, and therefore $a/a_\fpt=1$ and acceleration would be inefficient.

In analogy with the previous discussion for equatorial field lines,
let us now imagine uniformly expanding or contracting the field lines
near the axis.  That is, for each field line with a given
$\theta_\fpt\ll1$, let the polar angle far from the star become
$\theta=k\theta_\fpt$, with the same value of $k$ for all lines.  For
such a uniform expansion or contraction of the field, $B_p$ transforms
to $B_p/k^2$.  However, at the same time $R$ becomes $kR$, and so
$a=B_pR^2$ is unaffected.  In other words, there is no effect on
acceleration.

The key to obtaining acceleration along polar field lines is not
uniform lateral expansion (divergence) or contraction (collimation) of
field lines, but {\it differential bunching} of field lines.  To see
this rewrite equation (\ref{eq_gamma_minus_mu}) as
\begin{equation}
\frac{\gamma}{\mu} \approx1-\frac{a}{a_\fpt}
                   \approx 1 - \frac{\pi B_p R^2}{\Phi},
\end{equation}
where we have used the fact that $a_\fpt\approx\Phi/\pi$ near the pole.
Clearly, for efficient acceleration, we must make $B_p$ substantially
smaller than the mean enclosed field $\Phi/\pi R^2$.  That is, the
field lines interior to the reference field line must be bunched in
such a way that most of the flux has been pulled inward.  As an
example, consider a power-law distribution of the field strength,
\begin{equation}
B_p(R) \propto R^{-\xi},
\label{eq_Bpeta}
\end{equation}
where the index $\xi$ measures the degree of bunching.  This
distribution gives
\begin{equation}
\frac{\gamma}{\mu} \approx \frac{\xi}{2},
\label{eq_etaover2}
\end{equation}
which shows that the acceleration efficiency increases with increasing
$\xi$.  We reach equipartition between Poynting and matter energy
flux ($\sigma\sim1$, $\gamma\sim\mu/2$) for $\xi=1$, and we obtain
arbitrarily large efficiency ($\sigma\ll1$) as $\xi\to2$.

As we have described in \S\ref{sec_results}, polar field lines in simulation M90
are very efficient with $\gamma/\mu\to1$.  According to the above,
this would seem to suggest that $\xi(R)$ should be $\approx 2$ at the
axis and should decrease with increasing $R$.  The simulations,
however, show that this does not happen: as we show later,
$\xi\approx 1$, i.e., it stays roughly constant over a range of
$R$. Instead, higher efficiency near the jet axis is achieved in a
different way: as field lines bunch around the jet axis
\citep{chiueh_asymptotic_jet_structure_91,eichler_93,bogovalov_formation_jets_95,bog99}, they form a
concentrated core \citep{hn89,bog01,lyub_eichler_2001,bn09} that takes up a finite amount of flux
$\Phi_0$, leading roughly to the following poloidal field strength profile
\begin{equation}
B_p(R) = \Phi_0 \delta(\pi R^2) + B_0 (R/R_0)^{-\xi},
\label{eq_BpPhieta}
\end{equation}
where we have approximated the core profile with the Dirac delta-function.
This gives
\ifthenelse{\equal{\useiop}{0}}{
\begin{equation}
\frac{\gamma}{\mu}
        \approx \begin{cases}
          \simplefrac{\xi}{2}, \quad & \Phi_0 \ll \pi B_pR^2 / (1-\xi/2),\\
          1 - \simplefrac{\pi B_pR^2}{\Phi_0}, \quad & \Phi_0 \gg \pi B_pR^2 / (1-\xi/2).
        \end{cases}
        \label{eq_gammacore}
\end{equation}
}{
\begin{equation}
\frac{\gamma}{\mu}
        \approx \cases{
          \simplefrac{\xi}{2},  & $\Phi_0 \ll \pi B_pR^2 / (1-\xi/2)$, \cr
          1 - \simplefrac{\pi B_pR^2}{\Phi_0}, & $\Phi_0 \gg \pi B_pR^2 / (1-\xi/2)$.
        }
        \label{eq_gammacore}
\end{equation}
}
That is, in the limit when the flux in the concentrated core is small
compared to the flux in the surrounding power-law field distribution,
the efficiency is the same as in~\myeqref{eq_etaover2}. However, in the
opposite limit, i.e., sufficiently close to the core where the flux in
the core dominates, the angular profile of the poloidal field
distribution (and the value of $\xi$) become irrelevant for
determining the acceleration efficiency. Thus, (1) differential
bunching and (2) the resulting development of a concentrated core, are
the key requirements for efficient acceleration.

Note the following important corollary from the above discussion.  It
does not matter whether the particular field line of interest
collimates towards the axis or diverges from it. This has no effect on
the acceleration. What we need is that (1)~other field lines closer to
the axis must converge more, or diverge less, compared to the
reference field line, and/or (2)~a concentrated core at the jet axis
must contain a significant amount of magnetic flux.

\subsection{Comparison with Numerical Results}
\label{sec_comparison_with_numerical_results}

\begin{figure*}
\begin{center}
\includegraphics[width=0.8\textwidth]{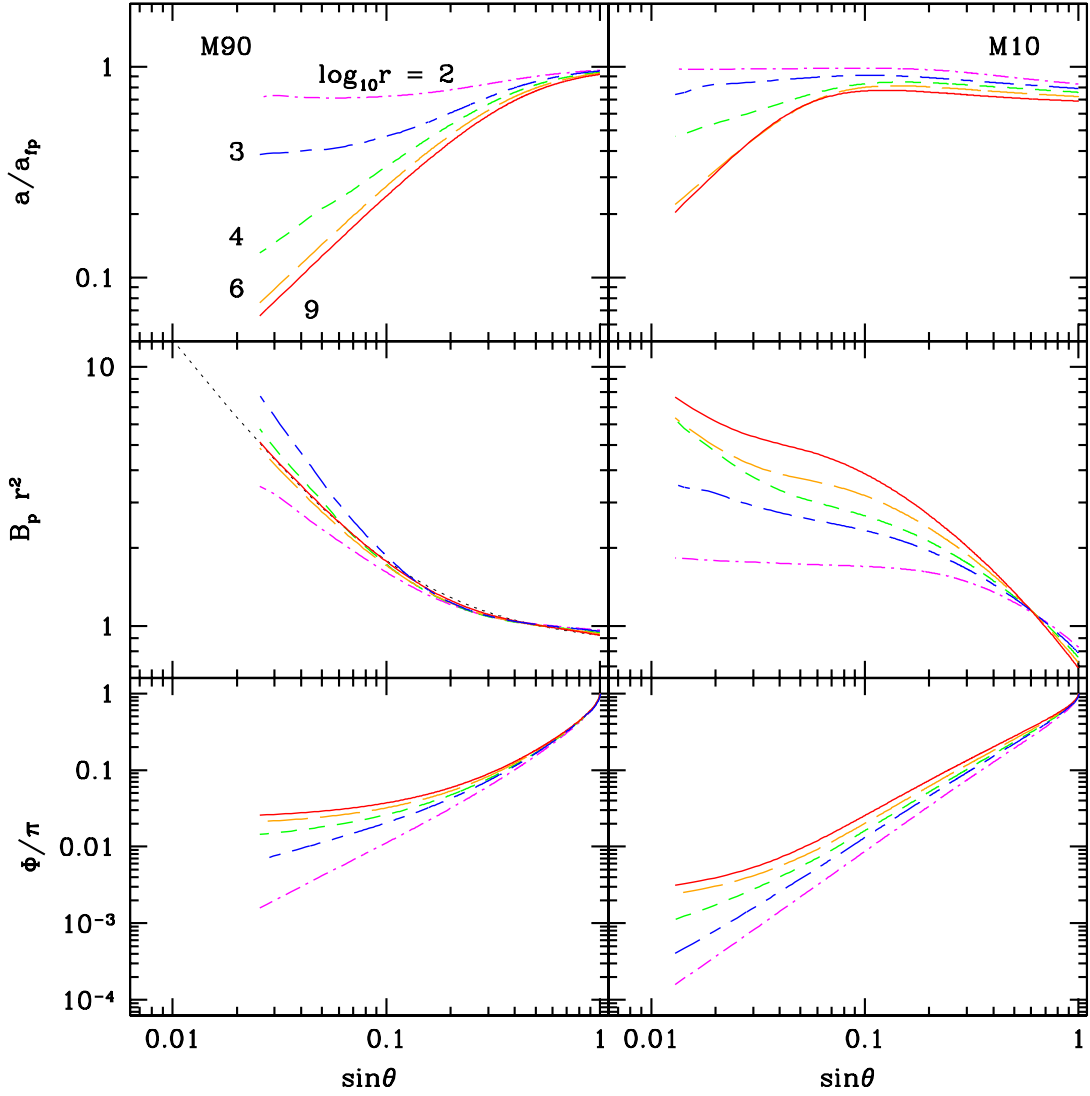}%
\end{center}
\caption{The panels on the left correspond to model M90 and those on
the right to M10.  The five curves correspond to different distances
from the star: $r=10^2$, $10^3$, $10^4$, $10^6$, $10^9$ (see Fig.~\ref{fig_2}
for a description of line colors and types).  The top panels show
the quantity $a/a_\fpt$, which determines the acceleration along a
field line.  Note that $a/a_\fpt$ is smallest near the poles,
where Fig.~\ref{fig_2} shows the largest acceleration.  The middle
panels show the profiles of normalized field strength $B_pr^2$ at the
same five distances.  Notice the nonuniform distribution as a function
of angle.  This is necessary for acceleration, as explained in
\S\ref{sec_analyt_bpnonuni}. In the middle-left panel the
curved dotted line corresponds to~\myeqref{eq_Bp_profile}
and accurately describes the angular profile of $B_p$ at
large distances.  The lower panels show the angular
distribution of the flux function $\Phi/\pi$.  }
\label{fig_aoafp}
\end{figure*}

Figure~\ref{fig_aoafp} shows results for models M90 and M10.  The top panels show
the behavior of $a/a_\fpt$ at different distances from the central
star.  We see that $a/a_\fpt$ decreases towards the pole, exactly
where $\gamma/\mu$ is largest and $\sigma$ is smallest in
Fig.~\ref{fig_2}.  Also, Fig.~\ref{fig_2} quantitatively confirms
the validity of equation (\ref{eq_gamma_minus_mu}) by plotting
prediction~\myeqref{eq_gamma_minus_mu} over the numerical solution.

The lower panels in Fig.~\ref{fig_aoafp} illustrate the effects described in the
previous two subsections.  In the equatorial regions, we see that
$B_pr^2$ decreases from its footpoint value of unity.  It is this
decrease that allows whatever acceleration is observed in this region
of the outflow.  However, the decrease is modest, so the acceleration
is not very large.

For angles closer to the pole, $B_pr^2$ actually increases relative to
its nominal initial value of unity. Nevertheless, this does not mean
that there is deceleration because, as we argued above, acceleration
near the axis is associated with differential bunching, not with any
overall expansion or contraction. For both M90 and M10, we see that
the magnetic field sets up the required bunching so that the poloidal
field is maximum at the axis and decreases with increasing distance
from the pole. It is this outward decrease, coupled with the presence
of magnetic flux $\Phi_0$ in a concentrated core at the pole (see
eq.~\ref{eq_gammacore}), that is associated with a decrease in
$a/a_\fpt$ for polar field lines and the reason for strong
acceleration.

The effect is most clearly seen in model M90, as illustrated in
Fig.~\ref{fig_aoafp}. Between $r = 10^2$ and $r = 10^3$ the angular
profile of $B_p$ becomes a steeper function of polar angle, leading to
an associated decrease in $a/a_\fpt$. Between $r=10^3$ and $r=10^4$
the trend reverses, and the poloidal field profile actually becomes a
shallower function of polar angle; however, $a/a_\fpt$ continues to
decrease. Beyond $r\sim10^4$ the magnetic field profile does not
evolve further.  In this regime it can be well-fitted by a broken power-law,
\begin{equation}
B_p r^2 = 0.85 + 0.075 \sin^{-1.1}\theta.
\label{eq_Bp_profile}
\end{equation}
Therefore at these distances $a=B_pr^2\sin^2\theta$ does not evolve
with $r$ either. Despite this, $a/a_\fpt\approx\pi a/\Phi$ decreases
with increasing $r$ and the value of $\Phi$ increases.  This is solely
due to the increase in the amount of magnetic flux $\Phi_0$ contained
in the concentrated core.

The picture that emerges is the following.  The poloidal magnetic field
establishes some equilibrium angular profile $B_p(r,\theta)r^2$ at
intermediate latitudes that does not evolve with $r$.  At
progressively larger $r$, each individual magnetic field line becomes
progressively more collimated towards the pole, but the same angular
profile of magnetic field is maintained.  This means that magnetic
flux $\Phi$ flows out from the low-latitude equatorial region ({}$B_p$
slightly decreases there), flows through intermediate latitudes
without changing the profile of $B_p$ there, and ends up in the
concentrated core at the polar axis, thereby uniformly shifting the
angular poloidal flux distribution $\Phi(\theta)$ up.  The effect is
clearly seen in the lower-left panel of Fig.~\ref{fig_aoafp}.

Note that it is not necessary to numerically resolve the concentrated
core in order to accurately describe the jet structure since it is
only the total amount of magnetic flux contained in the concentrated
core that matters for the acceleration efficiency. In particular, our
numerical method is well-suited for capturing such a core, even if
unresolved, since our method conserves the magnetic flux to machine
precision and can accurately capture the amount of magnetic flux that
enters the core and remains there.
We note that within a few grid cells from the polar axis, where the
unresolved magnetic flux accumulates, are least accurate
but this does not affect the quality of the solution at larger angles.
To verify this, we have checked
convergence of our models with angular resolution. For this, we ran a
version of model M90 that uses a uniform angular grid and has a factor
of $2$ lower effective angular resolution near the pole. Using this
less-resolved model leads to a maximum relative difference in the flux
function $\Phi$ and Lorentz factor of less than $15$\%, even near
the rotation axis. This difference is less than $2$\%
at most radii ($r<10^2$ and $r>10^5$) and is smaller at larger
$\theta$. This confirms the accuracy of the numerical solution.
Future higher-resolution models that resolve smaller angles and the
concentrated core are required to determine how the solution
connects the polar axis and for independent verification of results.

\section{Acceleration Efficiency and Communication with the Axis}
\label{sec_acceff}

\subsection{Fast Magnetosonic Surface}
\label{sec_fms}

\citet{bes98} have discussed the physical reason for inefficient
acceleration in the equatorial regions.  They show that it is related
to the fast magnetosonic point.  In the comoving frame of a cold MHD
plasma, fast magnetosonic waves travel with a speed $\mybeta_f$ given
by~\citep{gam03,mck06jf}
\begin{equation}
\gamma_f \mybeta_f = \left(\frac{b^2}{\rho}\right)^{1/2},
\end{equation}
where $b$ is the comoving magnetic field strength.
It is straightforward to relate $b$ to field components in the lab frame:
\begin{equation}
b^2 = B^2-E^2 \overset{\myeqref{eq_gammadrift}}\approx \frac{B^2}{\gamma^2}
              \overset{\myeqref{eq_eb_asymptotic}}\approx \frac{E \abs{B_\varphi}}{\gamma^2}
              \overset{\myeqref{eq_sigmadef}}\approx \rho\sigma.
\end{equation}
We thus find
\begin{equation}
\gamma_f \mybeta_f \approx \sigma^{1/2}.
\label{eq_betaf}
\end{equation}

Consider a streamline in the wind that moves outward with a local
Lorentz factor $\gamma$.  Let us first consider the limit of infinitely
high magnetization, \hbox{$\rho \to 0$}, \ie\
the force-free limit (\citealt{gol69,okamoto_magnetic_braking4_1974,
blandford_accretion_disk_electrodynamics_1976,lov76,bz77,macdonald_thorne_bh_forcefree_1982,
fendt_collimation1_1995,kom01,kom02,komiwaves02,mck06ffcode,nar07}; \citetalias{tchekhovskoy_ff_jets_2008}).
In this limit $\gamma_f\to\infty$, therefore
throughout the solution we have $\gamma<\gamma_f$.  This means that
the fast magnetosonic surface, defined by the condition
$\gamma=\gamma_f$, where the wind becomes causally detached from
fluid farther
back along its streamline, is located at infinity.
Such a force-free wind has a simple analytic solution in which
the Lorentz factor increases roughly linearly with distance~\citep{mic73},
\begin{equation}
\gamma \approx \Omega R. \label{eq_gamma_ff}
\end{equation}
In general, $\rho\ne0$, yet we might
expect that the behavior of the Lorentz factor in the sub-fast region
$\gamma<\gamma_f$ is similar to the force-free
solution~\myeqref{eq_gamma_ff}.  This has been shown to indeed be
the case~\citep{bes98}. However,
the acceleration in the super-fast
region $\gamma>\gamma_f$ has been found to become
logarithmic, \ie\ inefficient~\citep{bes98}.
Once the wind has crossed the fast magnetosonic point $\gamma=\gamma_f$,
it becomes causally detached from the fluid farther back along its
streamline.  We might therefore expect efficient acceleration
to cease beyond this fast magnetosonic
point\footnote{Note that fast magnetosonic waves move faster
than Alfv\'en waves, and so the causal horizon is determined by the
fast waves rather than Alfv\'en waves.} for all field lines.

If we define the fast Mach number $M_f$ by
\begin{equation}
M_f = \frac{\gamma\mybeta}{\gamma_f\mybeta_f}
    \approx \frac{\gamma\mybeta}{\sigma^{1/2}},
    \label{eq_Mfdef}
\end{equation}
then the fast point is the location at which $M_f\approx1$
(this equality would be exact if there was only motion
along the poloidal field line; however, there is also a slow
rotation in the toroidal direction which introduces a
negligible correction that we ignore). For a relativistic flow ($v\approx1$),
equation~\myeqref{eq_mu_via_gamma_sigma} lets us recast the above expression
in a useful form:
\begin{equation}
M_f^2 \approx \frac{\gamma^2}\sigma = \frac{\gamma^{3}}{\mu-\gamma}.
\label{eq_Mfsq}
\end{equation}

\begin{figure}
\begin{center}
\includegraphics[width=\columnwidth]{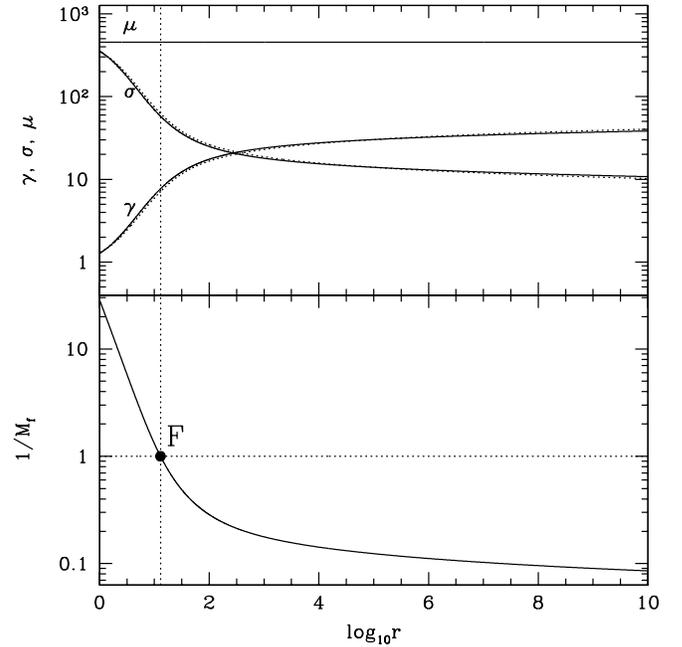}%
\end{center}
\caption{Dependence of various quantities as a function of distance
for a field line with $\theta_\fpt=\pi/2$ in model M90.  [Top panel]
Solid lines show the radial dependence of $\gamma$, $\sigma$, and
$\mu$ from the simulation. [Bottom panel] The solid line shows the
inverse of the fast Mach number $1/M_f$. The fast surface, indicated
with the vertical dotted line, is located where $M_f = 1$.  This is
also the location of the causality surface
\myeqref{eq_causality_surface} for this field line.  The various other
dotted lines show the analytical
approximation~\myeqref{eq_sigma_2asymptotes_from_causality}.}
\label{fig_along_m90}
\end{figure}

Figure~\ref{fig_along_m90} shows results for a field line in M90 with
$\theta_\fpt=\theta=\pi/2$.  Until the flow reaches the fast
magnetosonic point, we see that $\sigma$ falls rapidly and $\gamma$
increases rapidly.  However, both trends slow down substantially once
the flow crosses the fast point.  Beyond this point, \citet{bes98}
have shown that $\sigma$ and $\gamma$ vary as the one-third
power of $\log r$.  We confirm this dependence below.

The relatively abrupt cessation of acceleration beyond the fast point
for equatorial field lines is obvious in Fig.~\ref{fig_1}, where we
see that $\gamma$ stops increasing once the flow crosses the fast
magnetosonic surface (the middle of the three thick solid lines).
However, it is also clear from Fig.~\ref{fig_1} that something
else operates on polar field lines.  Model M10, in
particular, shows substantial continued acceleration well after polar
field lines have crossed the fast surface.  We discuss next the
relevant physics for these field lines.

\subsection{Communication with the Axis: Causality Surface}
\label{sec_causality}

We showed in \S\ref{sec_analyt} that, for efficient acceleration along a field line,
other neighboring field lines must shift laterally.  At the equator,
we need lines to move away from the midplane, while near the pole, we
need field lines to experience differential bunching or develop a flux core.
In order for any given field line to sustain efficient energy conversion,
it must be able to communicate to other regions of the
magnetized wind that undergo differential bunching or cause a
concentrated flux core.  This suggests
that the fast magnetosonic point, which determines where the fluid can
no longer communicate back along its motion, is perhaps not so
important.  A more relevant issue is whether or not the fluid can
communicate with regions near the axis that have field bunching or
a flux core.  We refer to the point at which a fluid
element loses contact with the axis as the ``causal point,'' and call
the locus of causal points over all field lines as the ``causality
surface''.  By the above arguments, we
expect that this surface, rather than the fast magnetosonic surface,
plays the role of the
boundary for efficient acceleration.\footnote{We note that 
the causality surface is formally different from the fast modified surface
\citep[which is discussed in detail in, e.g.,][]{guderley_1962,bp82,
contopoulos_plasma_gun_1995,tsinganos_critical_surfaces_1996}.
Both of these surfaces are built on an idea of a 
full causal disconnect: the causality surface requires a causal disconnect 
across the flow, while the fast modified 
surface requires a casual disconnect along the flow.}
We expect efficient acceleration inside this surface and
inefficient, logarithmic acceleration outside the surface~\citep[for
a related discussion, see][]{zbb08,komi_grb_jets_2008}.
In general fast waves propagate away from any given point toward the rotation
axis through an intermediate region where the density, velocity, and magnetic
field vary.
Hence, one should trace the position of fast waves emitted from any given point outward
over all angles and identify the ``causal point'' for each field line as where finally
no such traces can reach the polar axis.
For simplicity, we instead use only the local fast wave speed at a given point,
and we identify the approximate ``causal point'' by where the locally emitted
fast waves move away from any given point with a lab-frame local angle of $\theta>0$ with respect to the rotation
axis, so that the waves do not reach the rotation axis over a finite propagation distance.
We now calculate the approximate location of the causality surface using this
approach.

Consider a segment of the relativistic magnetized wind propagating
with a velocity vector $\myvec{\mybeta}$ and Lorentz factor $\gamma$ at an
angle $\theta_j$ to the rotation axis.\footnote{We
use $\theta$ for the polar coordinate of a point in
the solution and $\theta_j$ for the angle between the local poloidal
field and the axis.\label{ftn_thetaj}}
In \refapp{sec_mach_cone} we show that fast magnetosonic waves,
which are emitted isotropically in the comoving frame of the fluid, in the lab
frame will be collimated along $\myvec\mybeta$ into a Mach cone with a half-opening angle
\begin{equation}
\sin\xi_{\rm max} = \frac{\gamma_f \mybeta_f}{\gamma \mybeta} = \frac{1}{M_f}.
\label{ximax}
\end{equation}

By the argument given earlier, field line bunching and efficient
acceleration are possible only when the fluid can communicate with the
axis, i.e., only if $\xi_{\rm max} \gtrsim \theta_j$, i.e., only if
\begin{equation}
\sin\theta_j \lesssim \frac{1}{M_f}.
\label{eq_inside_c}
\end{equation}
For an equatorial wind, i.e., $\theta_j=\pi/2$, equation~\myeqref{eq_inside_c} shows
that acceleration stops when $M_f=1$, i.e., $\gamma=\gamma_f \approx
\sigma^{1/2}$ (\cf\ eq.~\ref{eq_betaf}, assuming $\mybeta_f\to1$).  That is, the wind stops
efficient acceleration as soon as it crosses the fast magnetosonic
point.  However, for smaller values of $\theta_j$, we obtain a
different result.

According to equation (\ref{ximax}), communication with the axis and
efficient acceleration are possible until
\begin{equation}
\gamma\mybeta \approx \frac{\gamma_f\mybeta_f}{\sin\theta_j} = 
\frac{\sigma^{1/2}}{\sin\theta_j}.
\label{eq_gammabeta}
\end{equation}
The presence of the factor~$\sin\theta_j$ in the denominator means
that communication extends to larger values of~$\gamma$, i.e.,
acceleration efficiency becomes larger as~$\theta_j$ decreases.  In
other words, polar field lines can accelerate more easily.
Using the definition of the fast magnetosonic Mach number~\myeqref{eq_Mfdef},
we obtain the following relation for the causality
surface:
\begin{equation}
\sin^2\theta_{j,c} \overset{\myeqref{eq_gammabeta}}{=}
             \frac{1}{M_{f,c}^2}
             \overset{\myeqref{eq_Mfsq}}{\approx}
             \frac{\sigma_c}{\gamma_c^2}
             = \frac{\mu-\gamma_c}{\gamma_c^3},
             \label{eq_causality_surface}
\end{equation}
where the subscript ``c'' indicates quantities evaluated at the
causality surface.
Inside the causality surface we expect the Lorentz factor to increase
roughly linearly with distance, \cf\ eq.~\myeqref{eq_gamma_ff}.  Based upon our earlier arguments,
once outside the causality surface the acceleration will only be logarithmic.
Using~\myeqref{eq_causality_surface}, we can estimate the distance at which
the causality surface is located:
\begin{equation}
r_c \sim \frac{\mu^{1/3}}{\Omega \sin^{5/3}\theta_{j,c}},
\label{eq_rc}
\end{equation}
where we have assumed that~\myeqref{eq_gamma_ff} and
$\gamma\ll\mu$ hold for $r \lesssim r_c$.

\begin{figure}
\begin{center}
\includegraphics[width=\columnwidth]{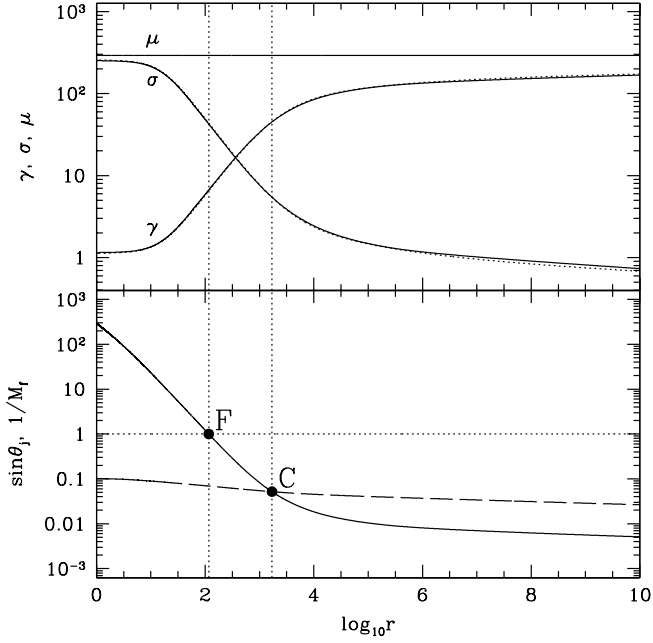}%
\end{center}
\caption{Dependence of various quantities as a function of distance
for a field line with $\sin\theta_\fpt = 0.1$ in model M10.  [Top panel]
Solid lines show the radial dependence of $\gamma$, $\sigma$, and
$\mu$.  [Bottom panel] The solid line shows the inverse of the fast
Mach number $1/M_f$ and the dashed line shows the local opening angle
of the field line $\theta_j$.  The two vertical dotted lines indicate
the positions of the fast magnetosonic point F ($M_f=1$) and the
causality point C ($\sin\theta_j = 1/M_f$).  The various other dotted
lines correspond to the analytical
approximation~\myeqref{eq_sigma_2asymptotes_from_causality}.}
\label{fig_along_m10}
\end{figure}

\begin{figure}
\begin{center}
\includegraphics[width=\columnwidth]{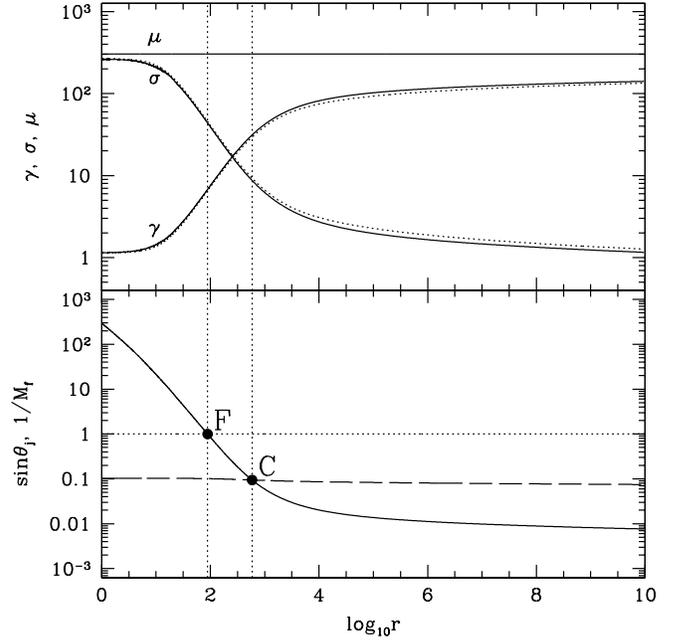}%
\end{center}
\caption{Dependence of various quantities as a function of distance
for a field line with $\sin\theta_\fpt = 0.1$ in model W10.  Comparison to
Fig.~\ref{fig_along_m10} shows that the wall inhibits the collimation of field
lines.  This brings the causality surface closer to the central star
than in the model without a wall, and hence leads to a lower
efficiency here compared to M10.  [Top panel] Solid lines show the
radial dependence of $\gamma$, $\sigma$, and $\mu$.  [Bottom panel]
The solid line shows the inverse of the fast Mach number $1/M_f$ and
the dashed line shows the local opening angle of the field line
$\theta_j$.  The two vertical dotted lines indicate the positions of
the fast magnetosonic point F ($M_f=1$) and the causality point C
($\sin\theta_j = 1/M_f$).  The various other dotted lines correspond
to the analytical
approximation~\myeqref{eq_sigma_2asymptotes_from_causality}.}
\label{fig_w10}
\end{figure}

\begin{figure}
\begin{center}
\includegraphics[width=\columnwidth]{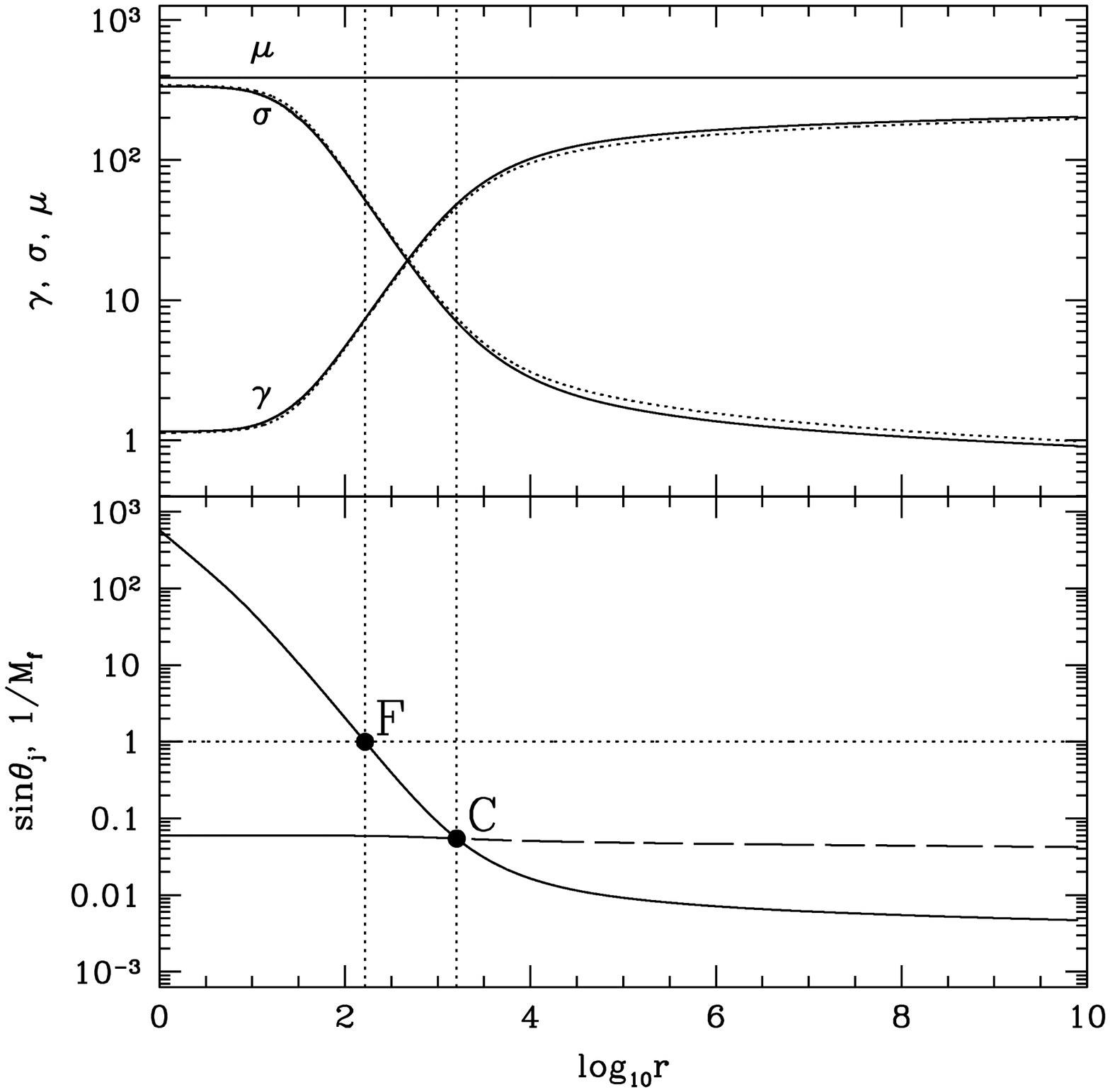}%
\end{center}
\caption{Dependence of various quantities as a function of distance
for a field line with $\sin\theta_\fpt = 0.06$ in model W5.  [Top panel]
Solid lines show the radial dependence of $\gamma$, $\sigma$, and
$\mu$.  [Bottom panel] The solid line shows the inverse of the fast
Mach number $1/M_f$ and the dashed line shows the local opening angle
of the field line $\theta_j$.  The two vertical dotted lines indicate
the positions of the fast magnetosonic point F ($M_f=1$) and the
causality point C ($\sin\theta_j = 1/M_f$).  The various other dotted
lines correspond to the analytical
approximation~\myeqref{eq_sigma_2asymptotes_from_causality}.}
\label{fig_w5}
\end{figure}

Figure~\ref{fig_1} confirms that the causality surface (the outermost of the
three thick solid lines) provides a better approximation to the
boundary between the acceleration and coasting zones compared to the
fast magnetosonic surface.  Figure~\ref{fig_along_m10} shows detailed results for a
polar field line with $\sin\theta_\fpt=0.1$ in model M10.  Notice that
efficient acceleration continues well past the fast magnetosonic point
(``F'', dotted line on the left); acceleration slows down only after the
field line has crossed the causal point (``C'', dotted line on the right).
This is the reason why this particular field line is able to achieve a
Lorentz factor of $170$ with a high efficiency of
$\gamma/\mu\approx0.6$, which is much larger than for equatorial field lines.

\subsection{Analytical Approximation}
\label{sec_analyt_approx}

We now develop an analytical approximation to calculate the Lorentz
factor as a function of distance for any field line in a monopole
magnetized wind.  Generally, for a force-free jet, there exist two
distinct acceleration regimes, as explained in~\citetalias{tchekhovskoy_ff_jets_2008}.
In \refapp{sec_gamma_mhd}
we generalize these results to MHD (finite-magnetization) jets.  We
summarize the results here.  In the first
acceleration regime, which is realized near the
compact object, the Lorentz factor of the flow increases roughly
linearly with distance:
\begin{equation}
\gamma_1^2 \approx \gamma_0^2 + (\Omega R)^2,
\label{eq_sigma1}
\end{equation}
where $\gamma_0$ is the initial Lorentz factor at $r=1$.

Based on earlier arguments, beyond the causality surface the acceleration is only logarithmic.
This is the second acceleration regime in which the Lorentz factor
is determined by the poloidal shape of the field lines~(\citealt{bes98}; \citetalias{tchekhovskoy_ff_jets_2008}).
Using the results of \refapp{sec_logthird} as a guide \citep[see also][]{bes98,
lyub_eichler_2001}, we expect in this region
\begin{equation}
\gamma_2 \propto \ln^{1/3}r.  \label{eq_gammaalog}
\end{equation}
To make this formula quantitative, we demand that
it gives the correct value of the Lorentz factor
at the causality point $r = r_c$ (see eq.~\ref{eq_causality_surface}):
\begin{equation}
\gamma_c \approx \mu^{1/3} \sin^{-2/3}\theta_{j,c}. \nonumber
\end{equation}
We do this by choosing the solution in the following form:
\begin{eqnarray}
\gamma_2 &\approx&
         C_1 \, \gamma_c \, \ln^{1/3}(1+C_2r/r_c) \\
         &\approx&
         C_1 \, \mu^{1/3} \frac{\ln^{1/3}(1+C_2r/r_c)}{\sin^{2/3}\theta_{j,c}},
\label{eq_sigma2gg1}
\end{eqnarray}
where $C_1$ and $C_2$ are
numerical factors of order unity that we later determine by fitting
to the numerical solution.  The radial and angular
scalings in equation~\myeqref{eq_sigma2gg1}
agree with the analytic expectations~\citep{bes98,lyub_eichler_2001,lyub09}.

Note, however, that formulae~\myeqref{eq_gammaalog}--\myeqref{eq_sigma2gg1}
become inconsistent at low magnetization since $\gamma$
cannot exceed $\mu$, whereas the right-hand sides of these equations are
unbound.  Noting that the fast wave Mach number $M_f$ is unbound and
$\gamma/\mu^{1/3} \approx M_f^{2/3}$ for $\gamma\ll\mu$ (eq.~\ref{eq_Mfsq}),
we empirically
modify~\myeqref{eq_sigma2gg1} by replacing $\gamma_2/\mu^{1/3}$ with $M_f^{2/3}$:
\begin{equation}
M_f^{2/3} = C_1 \frac{\ln^{1/3}(1+C_2r/r_c)}{\sin^{2/3}\theta_{j,c}},  \label{eq_Mflog}
\end{equation}
where $C_1$ and $C_2$ are numerical factors of order unity (see below).
Substituting for the fast wave Mach number
$M_f$ using equation~\myeqref{eq_Mfsq},
we obtain a cubic equation for the
Lorentz factor on the field line beyond
the causality surface, $r\gtrsim r_c$:
\begin{equation}
\gamma_2 = C_1 \, (\mu-\gamma_2)^{1/3}\frac{\ln^{1/3}(1+C_2 r/r_c)}{\sin^{2/3}\theta_{j,c}},
\label{eq_sigma2}
\end{equation}
where $\mu$ is the value of the total specific energy flux on the
field line in question (eq.~\ref{eq_mudef}),
$\theta_{j,c}$ is the value of the angle~$\theta_j$ that the
field lines makes with the polar axis at the causality surface
(see footnote~\ref{ftn_thetaj}), and $C_1 \simeq C_2 \simeq 1$ (see below).
In the limit $\gamma_2\ll\mu$ this equation reduces to~\myeqref{eq_sigma2gg1}.

We now combine the two approximations~(\ref{eq_sigma1}) and
(\ref{eq_sigma2}) to write (see \refapp{sec_gamma_mhd})
\begin{equation}
\frac{1}{\gamma^2} = \frac{1}{\gamma_1^2} +\frac{1}{\gamma_2^2}.
\label{eq_sigma_2asymptotes_from_causality}
\end{equation}
Clearly, the \emph{smaller} of $\gamma_1$ and $\gamma_2$
determines the total Lorentz factor: in accordance with the above
discussion, near the compact object
$\gamma\approx\gamma_1$ and at a large distance (outside the causality
surface)  $\gamma\approx\gamma_2$.
We find that we obtain good agreement with our simulation results when
we choose $C_1=2$, $C_2 = 0.4$.  In fact, for this single
set of parameters formula~\myeqref{eq_sigma_2asymptotes_from_causality},
with $\gamma_1$ and $\gamma_2$ given by~\myeqref{eq_sigma1} and \myeqref{eq_sigma2},
does quite well
for all field lines in all simulations, both in the limit of low and
high magnetizations.  The various dotted lines in
Figs.~\ref{fig_along_m90}--\ref{fig_transversalall} have
all been calculated using this formula, and clearly provide an
excellent representation of the numerical results.

\subsection{Other Models}
\label{sec_comparison}

We have so far discussed in detail the representative models M90 and
M10. However, we have carried out a number of other simulations. We
mentioned models M20 and M45 in \S\ref{subsec_M10}. We have also
carried out models with walls: W5, W10.

Simulation W10 has the same setup as M10 but has an impenetrable
perfectly conducting wall at $\theta = 10^\circ$. Figure~\ref{fig_w10}
shows that the wall in simulation W10 keeps field lines near the wall
from collapsing onto the pole and prevents the rest of the field lines
from developing the lateral nonuniformity
required for efficient acceleration.  As a result, field lines very close
to the wall accelerate more efficiently but the rest of the field lines
have a suppressed efficiency: a field line with
$\sin\theta_\fpt = 0.1$ in model W10 has a lower Lorentz factor at $r
= 10^{10}$, $\gamma \approx 150$, than the corresponding field line in
model M10, which has $\gamma\approx 170$ (compare Figs.~\ref{fig_along_m10}
and~\ref{fig_w10}).  Simulation W5, which has the wall at $\theta =
5^\circ$, is the most collimated model that we have
simulated. Figure~\ref{fig_w5} shows a field line for that model that
makes an angle $\sin\theta_\fpt = 0.06$ at the surface of the central
star. This field line reaches equipartition by $r \sim 10^{10}$ with
$\gamma \approx 200$, the largest Lorentz factor we have achieved
among all simulations in this paper.

\begin{figure*}
\begin{center}
\includegraphics[width=0.49\textwidth]{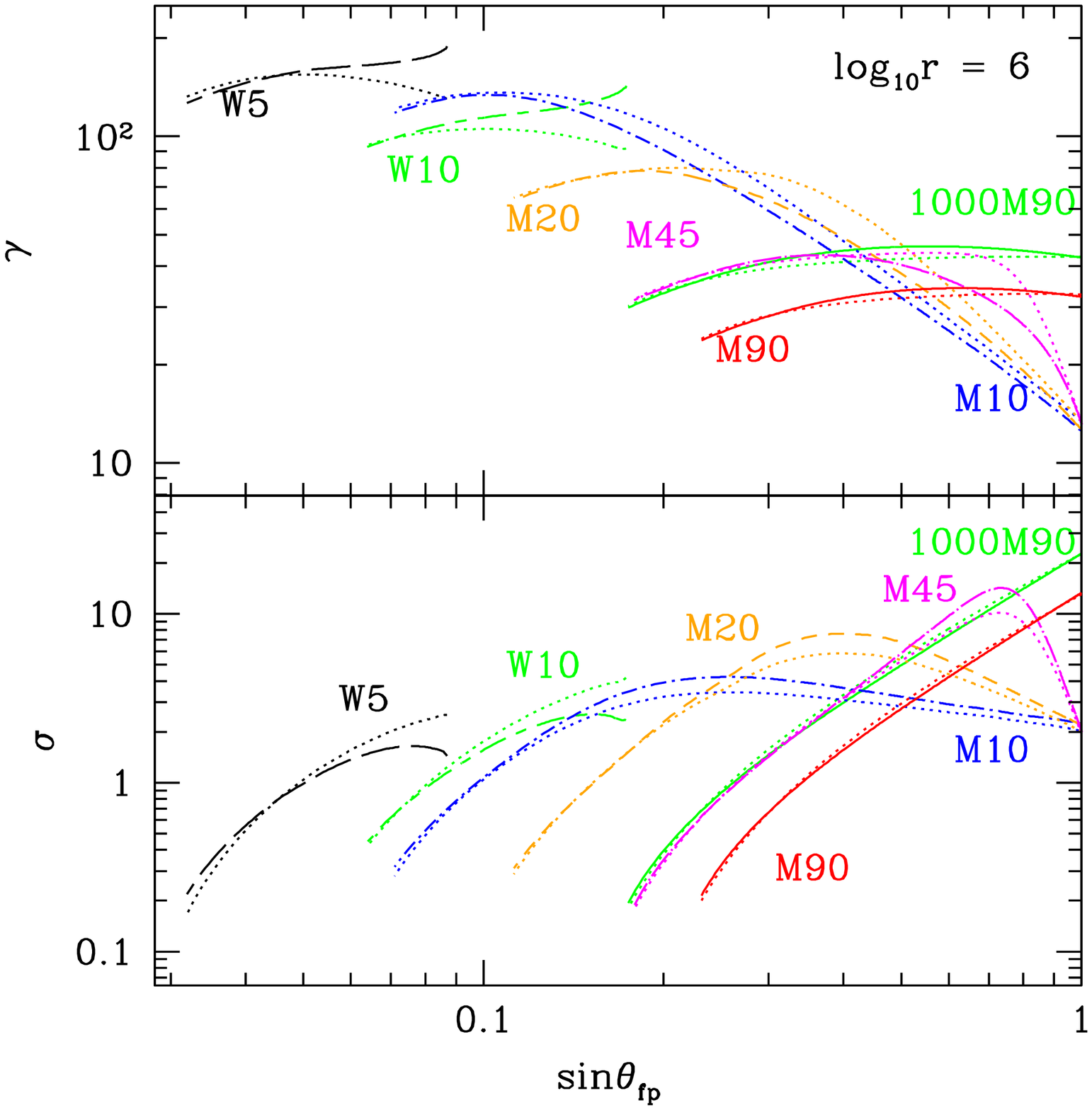}%
\hfill\includegraphics[width=0.49\textwidth]{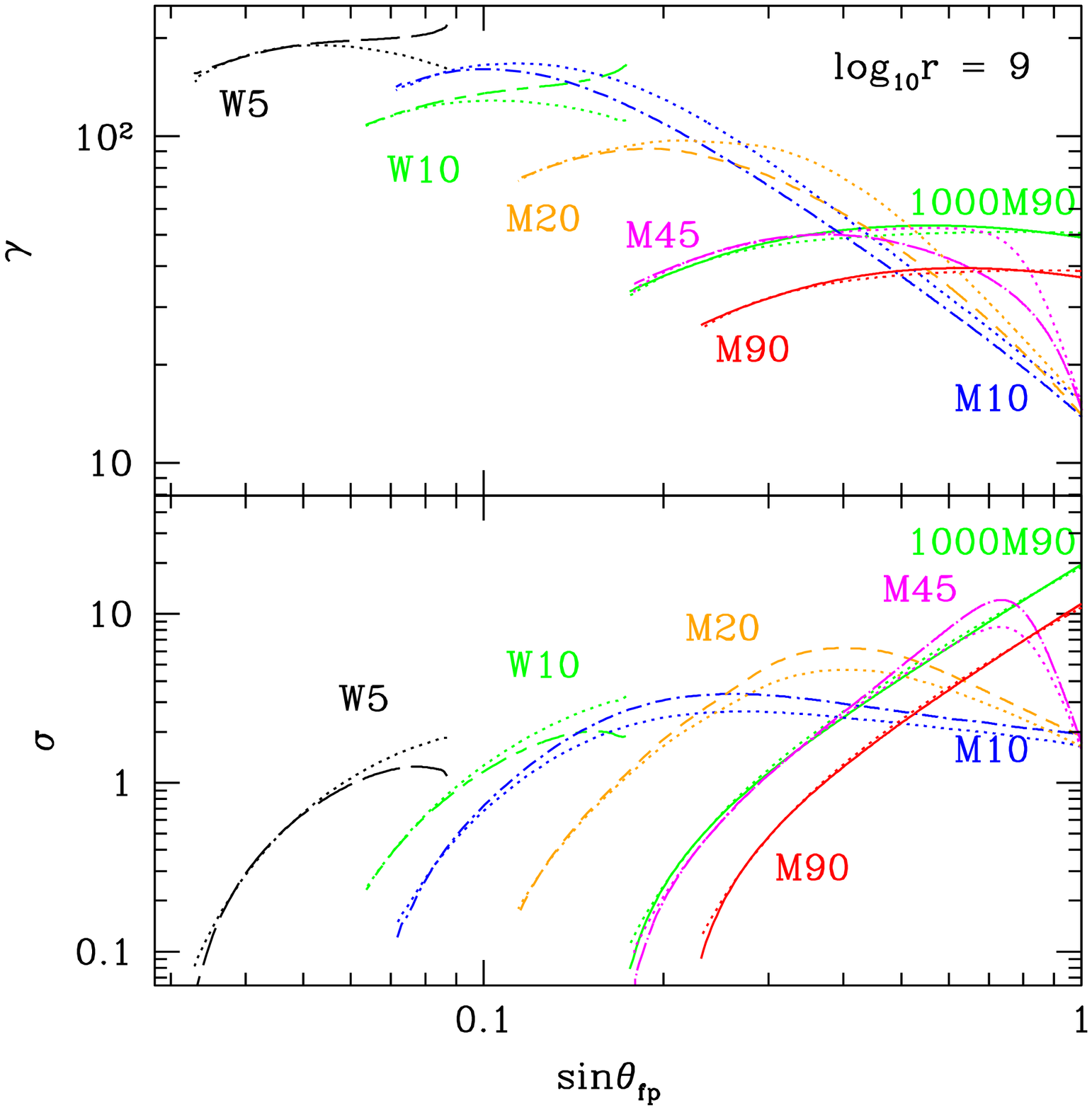}%
\end{center}
\caption{Dependence of $\gamma$ (upper panels) and $\sigma$ (lower
panels) as a function of $\sin\theta_\fpt$ at $r=10^6$ (left panels)
and $r = 10^9$ (right panels) for a series of models. The names of the
models are indicated next to the corresponding curves.  Dotted lines
show the analytic solution~\myeqref{eq_sigma_2asymptotes_from_causality}
and other lines show the simulation results.  For models with a
constant density profile $\rho(\theta_\fpt)$ on the star (M90, 1000M90), the analytic
approximation works very well at all distances $r$ and all polar
angles. For models with a variable angular profile of density on the
surface of the star (M45, M20, M10), the agreement is excellent along the field lines
for which the initial density is constant; for other field lines, the
agreement is less perfect, indicating the best-fit values of the
factor $C_1$ in equation~\myeqref{eq_sigma2} are different for different
field lines. For models with the wall (W10, W5), the agreement is
good near the axis and poor near the wall.}
\label{fig_transversalall}
\end{figure*}

We have also performed a simulation called 1000M90 that has a uniform
density profile on the stellar surface $\rho(\theta_\fpt) = \rho_0$
with the same density at $\theta_\fpt=0$ as model M45.  Thus,
models 1000M90 and M45 have similar profiles of density and $\mu$ near the
pole, but they differ near the equator.  We expect the two simulations
to show nearly identical behavior for polar field lines.  This is
indeed confirmed, as seen in Fig.~\ref{fig_transversalall}. The point
of this model is to verify that models with variable density, e.g.,
equation~\myeqref{eq_bc_rho}, give reliable results near the axis, independent of how
we modify the mass-loading of equatorial field lines.  The numerical
results confirm that this is indeed so.

Figure~\ref{fig_transversalall} shows transversal cuts through each of
our models at distances of $10^6$ and $10^9$, and compares the
simulation results with the analytic
approximation~\myeqref{eq_sigma_2asymptotes_from_causality} described
in~\S\ref{sec_analyt_approx}. For models with a constant density profile
$\rho(\theta_\fpt)$ on the star, the analytic approximation works
extremely well at all distances $r$ and all polar angles. For models
with a variable angular profile of density on the surface of the star,
the agreement is excellent along the field lines originating in the
constant-density core while for other field lines, the best-fit values
of factor $C_1$ in equation~\myeqref{eq_sigma2} apparently varies from
one field line to the next. Therefore, adopting a single value $C_1 =
2$ provides only a rough description of acceleration along these field
lines.

\section{Discussion}
\label{sec_discussion}

A number of previous authors have noted that efficient acceleration of a cold
MHD wind requires field lines to diverge away from the equatorial plane.
This field geometry was identified as a ``magnetic nozzle''
due to the geometric similarity of jet nozzles intended to launch a supersonic flow.

We generalized the concept of the ``magnetic nozzle'' by
showing that the geometric bunching of field lines
generally induces efficient conversion of magnetic to kinetic energy in neighboring regions.
We clarified how this generalized ``magnetic nozzle'' operates differently
for equatorial and polar field lines.
Near the midplane, lines merely have to diverge uniformly away from the equator.
The more they diverge, the larger the acceleration.  For polar field
lines, however, what is needed is neither simple divergence nor
convergence, but differential bunching or a region that accumulates flux.
A particular field line may either converge or diverge relative to its initial (purely radial)
configuration.  This has no effect on acceleration along this line.
However, if neighboring field lines move such that the field strength
decreases away from the rotation axis, e.g., as per the simple
prescription given in equation~\myeqref{eq_BpPhieta}, then acceleration
will occur along the reference field line.  The more the differential
bunching (i.e., the larger the value of $\xi$ or $\Phi_0$), the larger
the acceleration.

The rearrangement of field lines described above requires different
regions of the magnetosphere to communicate with one another, which is
possible only if the flow speed is not too large.  This introduces the
second major difference between equatorial and polar field lines.
Equatorial field lines lose communication once their flow velocities
cross the fast magnetosonic speed.  This happens when the Lorentz
factor $\gamma\sim\mu^{1/3}$, where $\mu$ is the conserved energy flux
per unit mass flux along the line.  Equivalently,
$\gamma\sim\sigma^{1/2}$, where $\sigma$ is the local magnetization
parameter.  For highly relativistic flows, $\mu$ is very large (e.g.,
$\mu\sim10^6$ for the Crab Pulsar), so $\gamma$ is much less than
$\mu$ at the fast magnetosonic transition.  Correspondingly,
$\sigma\sim\mu^{2/3}$ is very large, which means that most of the
energy flux is still carried as Poynting flux rather than as mass
energy flux.

For equatorial field
lines beyond the fast point, a small amount of further acceleration is
possible, but this only gives an additional logarithmic factor
\citep{bes98}. After allowing for this factor, the final asymptotic
Lorentz factor on an equatorial field line at a large distance from
the star is only $\gamma_\mathrm{asym}\sim 2\mu^{1/3}\ln^{1/3}(\Omega
r/\mu^{1/3})$ (see eqs.~\ref{eq_rc} and \ref{eq_sigma2gg1}). This is far smaller than the maximum Lorentz factor one
would obtain if we had efficient acceleration along the field line, viz.,
$\gamma_{\rm max}=\mu$. Thus, we confirm the previously known result
that equatorial field lines in monopole geometry suffer from a serious
$\sigma$ problem. We do not yet see any way of avoiding this
conclusion.

The situation is different for polar field lines.  Even
beyond the fast magnetosonic point, the fluid on these field lines can maintain communication
with the axis (where field bunching allows efficient
energy conversion).  In fact, communication is lost only when
$\gamma\sim\sigma^{1/2}/\sin\theta_j$, where $\theta_j$ is the angle
between the poloidal component of the magnetic field and the rotation
axis.  For small values of $\theta_j\sim0.01-0.1$, as might be
appropriate for relativistic jets, this gives a large increase in the
asymptotic Lorentz factor reached by the flow.  Including the
additional gain from the logarithmic factor,
we estimate~(\cf\ eqs.~\ref{eq_rc} and~\ref{eq_sigma2})
\begin{equation}
\gamma_\mathrm{asym}\approx (\mu-\gamma_\mathrm{asym})^{1/3}\frac{C_1\ln^{1/3}(\Omega r\sin^{5/3}\theta_{j,c}/\mu^{1/3})}{\sin^{2/3}\theta_{j,c}},
\label{eq_gamma_via_mu_sin}
\end{equation}
or, using~\myeqref{eq_mu_via_gamma_sigma},
\begin{equation}
\gamma_\mathrm{asym}\approx \sigma^{1/2}\frac{C_1^{3/2}\ln^{1/2}(\Omega r\sin^{5/3}\theta_{j,c}/\mu^{1/3})}{\sin\theta_{j,c}},
\label{eq_gamma_via_sigma_sin}
\end{equation}
where we note that $\mu-\gamma_\mathrm{asym}\propto\sigma\propto\sin^2\theta_{j,c}$
near the rotation axis.
In the limit $\sigma\gg 1$ we can simplify~\myeqref{eq_gamma_via_mu_sin} by approximating
\mbox{$\mu-\gamma_\mathrm{asym}\approx \mu$}. These expressions are
applicable beyond the causality surface as approximately given by equation~\myeqref{eq_rc}.
As an order of magnitude estimate, in these formulae one could substitute $\theta_\fpt$, $\theta$,
or $\theta_j$ in place of $\theta_{j,c}$. In any case, the $\sin\theta_{j,c}$
factors in the denominator of equations~\myeqref{eq_gamma_via_sigma_sin}
and~\myeqref{eq_gamma_via_mu_sin} indicate that $\gamma_\mathrm{asym}$ is larger
near the poles.

\subsection{Collimation and Acceleration}
\label{sec_collimation_and_acceleration}

Equations~\myeqref{eq_gammabeta}, \myeqref{eq_gamma_via_sigma_sin} predict that at roughly the
same value of $\sigma\sim1$, an asymptotically more collimated
simulation reaches a larger Lorentz factor:
\begin{equation}
\gamma_\mathrm{asym}\propto\frac{1}{\sin\theta_{j,c}},
\label{eq_gamma_inverse_thetajc}
\end{equation}
where $\theta_{j,c}\sim\theta_\fpt\gtrsim\theta_j$ is the angle at the causality surface.
Figure~\ref{fig_transversalall} confirms this for the sequence of models
M45--M20--M10, all of which have the same maximum value of $\mu$,
$\mu_{\rm max} = 460$.  If all of these models could convert all of
the electromagnetic energy flux into kinetic energy flux, each of them would
reach the maximum energetically allowed Lorentz factor, $\gamma = \mu_{\rm max}$.
While such a full conversion does not happen in any of these models,
more collimated models reach higher Lorentz factors, in agreement
with equation \myeqref{eq_gamma_inverse_thetajc}:
each subsequent model in the sequence of models M45--M20--M10
is roughly twice as collimated as the previous one and at $r=10^9$ attains
approximately twice as large a Lorentz factor, $50$--$95$--$170$.

\subsection{Application to Relativistic Jets and GRBs}

The net conclusion of the previous discussion is that, whereas there
is indeed a serious $\sigma$ problem for pulsar winds, there is no
similar problem for relativistic jets.  We show that even if a flow is
unconfined and has a monopolar-like shape, it can still efficiently
accelerate in the polar region.  For a jet
angle $\theta_j\sim2^\circ$, for instance, the scaling~\myeqref{eq_gamma_via_sigma_sin}
predicts an asymptotic Lorentz factor
$\gamma_\mathrm{asym} > 100$ if the jet efficiently converts
electromagnetic to kinetic energy and reaches
$\sigma\sim1$.  In fact, if the jet is not efficient and carries more
of its energy as Poynting flux (\eg\ for a larger value of $\mu$, see
eqs.~\ref{eq_gamma_via_sigma_sin}--\ref{eq_gamma_via_mu_sin}), then $\sigma>1$, and we will have even
larger values of $\gamma_\mathrm{asym}$.  These estimates are confirmed by
the numerical simulations described in this paper.

The inferred total power of long GRBs is on the order of $10^{51}$~erg~\citep{pir05,meszaros_grbs_2006,liang2008};
however, much less energetic events, with total energy release as
low as $10^{48}$~erg, have also been observed \citep{sod_subenergetic_04,sod_subenergetic_06}.  Let us compute the
power output of jets in our numerical models and make sure that our jets
are energetic enough to be consistent with these observations.
The total power coming out from the compact object surface within the
low-density core, $\theta < \theta_{\rm max}$, is
\begin{eqnarray}
P^{\rm jet} &=& \int\limits_0^{\theta_{\rm max}} 2\pi r^2d\theta\sin\theta \, S(r,\theta)|_{r=1}
       \approx \frac{\Omega^2}{2}\int\limits_0^{\theta_{\rm max}} d\theta\sin^3\theta \nonumber\\
       &=& \frac{\Omega^2}{2} \times \frac{4(2+\cos\theta_{\rm max})}{3}
          \times \sin^4\frac{\theta_{\rm max}}{2}
       \approx  \frac{\Omega^2}{8} \,\theta_{\rm max}^4,
\end{eqnarray}
where the last equality is for $\theta_{\rm max}\ll 1$.
Converting the result to physical units, we obtain
\begin{equation}
P^{\rm jet} \approx \frac{1}{8}\Omega^2 B_r^2 r_0^4 \thetamax^4/ c,
\label{eq_jet_en_frac}
\end{equation}
where $B_r$ is the value of radial magnetic field component on
the surface of the compact object and $r_0$ is the compact object radius.
Evaluating the jet power for a magnetar with a characteristic
period $P = 1$~ms and a surface magnetic field of $10^{15}$~G, we get
\begin{eqnarray}
P^{\rm jet}_{\rm NS} &\approx& 2 \times 10^{47}  \left[\frac{\mathrm{erg}}{\mathrm{s}}\right]\nonumber\\
                &\times&\left(\frac{1 \mathrm{ms}}{P}\right)^2
                \left(\frac{B_r}{10^{15} \mathrm{G}}\right)^2
                \left(\frac{r_0}{10 \mathrm{km}}\right)^2
                \left(\frac{\thetamax}{10^\circ}\right)^4.
                \label{eq_grb_energy_ns}
\end{eqnarray}
For a maximally-spinning black hole with dimensionless spin parameter
$a = 1$, mass $3M_\odot$,
and surface magnetic field strength
$10^{16}$~G \citep{mck05}, we have
\begin{equation}
P^{\rm jet}_{\rm BH}  \approx 5 \times 10^{48}
                \, a^2
                \left(\frac{B_r}{10^{16} \mathrm{G}}\right)^2
                \left(\frac{M}{3M_\odot}\right)^2
                \left(\frac{\thetamax}{10^\circ}\right)^4
                \left[\frac{\mathrm{erg}}{\mathrm{s}}\right].
                \label{eq_grb_energy_bh}
\end{equation}
Given a characteristic duration of $10$--$100$ seconds for a long GRB,
our numerical jets provide, for the black hole case,
$10^{50}$--$10^{51}$~erg per event for model
M10, $10^{51}$--$10^{52}$~erg for M20,
and $10^{52}$--$10^{53}$~erg for M45.
Therefore, the simulated jets from black holes in models
M10 and M20 are energetic enough and move at sufficiently
high Lorentz factors ($\gamma\gtrsim100$) to account for most long
GRBs and can certainly account for low luminosity events.
The energetics is also right for short GRBs:
the simulated jets output $10^{49}$--$10^{50}$ erg
during a characteristic event duration of $1$~second~\citep{nakar_short_grbs_2007}.
For the magnetar case the energetics
is lower: $10^{47}$--$10^{48}$--$10^{50}$~erg/s for a
sequence of models M10--M20--M45.
Therefore, the simulated jets from magnetars can account for
less-luminous long GRB events and most short GRBs.

While the energy fraction in the polar jet  is small as compared to the
total energy extracted by magnetic fields from the black hole,
$P^{\rm jet}/P^{\rm tot} \sim\theta_{\rm max}^4$ (eq.~\ref{eq_jet_en_frac},
which assumes a uniform magnetic field distribution at the BH horizon), the
absolute value of jet power $P^{\rm jet}$ is sufficiently large to account for long
and short GRBs. We point out that the magnetic flux in accreting black
hole systems is non-uniformly concentrated in the polar region of the
BH \cite[\eg,][due to ambient pressure of the accretion flow]{mck05}, and therefore the
total energy losses of the spinning black hole are actually dominated
by the losses from the polar region rather than from the midplane
region, meaning a larger fraction of power in the jet than given by
eq. \myeqref{eq_jet_en_frac}.

A very interesting question is whether the models suggest any
characteristic value for the quantity $\gamma_j\theta_j$: is this
quantity generally smaller or larger than unity? For a jet with
$\gamma_j\theta_j\gg1$ (say, $\sim10$) only part of the jet within the
beaming angle $\theta_b\approx1/\gamma_j \ll \theta_j$ is visible to a
remote observer. As the interaction with the ambient medium
decelerates the jet and the beaming angle becomes comparable to the
jet opening angle, the edges of the jet come into sight and the light
curve steepens achromatically, displaying a ``jet break'' \citep{pir05,
 meszaros_grbs_2006}. In the other limit, $\gamma_j\theta_j\ll1$, a jet
would be incapable of producing achromatic breaks in GRB light curves.

Achromatic breaks have been found in pre-\Swift\
times~\citep{frail01,bfk03,zeh_long_grb_angles_2006} and have yielded
$\theta_j\sim0.03$--$1$ rad.  The situation is quite different
in the post-\Swift\ times for which a large amount of data available, and
no jet breaks have been found to fully satisfy closure relations
in all bands \citep{liang2008}.  However, if one or more of the closure relations
are relaxed, some of the breaks may be interpreted as ``achromatic''
and be used to derive jet opening angles that span a similar range as
pre-\Swift\ GRBs \citep{liang2008}.
Overall, it appears that the Lorentz factor of
most GRBs is $\gamma_j\gtrsim100$ (\citealt{pir05,meszaros_grbs_2006},
up to $\sim400$, \citealt{lithwick_lower_limits_2001}) which, with
the above estimates for $\theta_j$, gives
$\gamma_j\theta_j \gtrsim 3$ and indicates that in principle
achromatic jet breaks are possible.

Our numerical models have $\gamma_j\theta_j\sim 5$--$15$, where the
index $j$ denotes quantities evaluated at the jet boundary which
we define as the boundary of matter-dominated region $\sigma<1$. In this
respect it is particularly fruitful to compare simulations M10 and
W10. As we discussed in~\S\ref{sec_comparison}, the wall in model W10
prevents field lines from collapsing onto the pole as much as they do
in model M10. Comparison of field lines with $\sigma\sim1$ for these
models (see Figs.~\ref{fig_along_m10} and \ref{fig_w10}), reveals a
difference in the Lorentz factor of at most $20$\% and a much larger
difference in the value of $\gamma_j\theta_j$ (caused by a large
difference in $\theta_j$ due to the effect of the wall):
$\gamma_j\theta_j\approx 5$ for M10 and $\approx 12$ for W10. The most
collimated model W5 produces an even larger value, $\gamma_j\theta_j
\approx 15$.

We now analytically confirm that in
general the quantity $\gamma_j\theta_j \sim 5$--$10$.
For the jet boundary, using~\myeqref{eq_gamma_via_sigma_sin} and
characteristic values $\Omega \sim 1$, $\theta_{j,c} \sim 0.04$--$0.4$,
$\mu\sim10^3$, $r\sim10^6$--$10^9$, $\theta_j/\theta_{j,c}\sim 0.5$--$1$ (see
Figs.~\ref{fig_along_m10}, \ref{fig_w10}, and \ref{fig_w5}),
we get:
\begin{equation}
\gamma_j\theta_j \sim \frac{4\theta_j}{\theta_{j,c}} \sigma^{1/2}\log_{10}^{1/2} \left(\frac{r}{10^2{\!-\!}10^3}\right)
               \sim (4{\!-\!}10)\sigma^{1/2}
\label{eq_gammathetajnum}
\end{equation}
which is in good agreement with the simulation results.
The large value of $\gamma_j\theta_j$ in this analysis
arises solely due to the logarithmic factor that appears because
a significant fraction of the acceleration
occurs after crossing the causality surface, in the
inefficient acceleration region:
this is the case for all unconfined flows studied in our paper.
This should be contrasted with confined outflows, collimated by
walls with prescribed shapes, for which
most of the acceleration tends to complete before crossing
the causality surface and which have
$\gamma_j\theta_j\sim 1$ \citep{komi_grb_jets_2008}.
We note that if we do not require that jets are matter-dominated, \eg\
we allow $\sigma>1$ as in \citeauthor{lyutikov_grbs_2003}'s
(\citeyear{lyutikov_grbs_2003}) model of GRBs,
then the value of $\gamma_j\theta_j$ will be even higher (see
eq.~\ref{eq_gammathetajnum}).

According to equation~\myeqref{eq_gammathetajnum}, we can attribute the
fact that some post-\Swift\ GRBs show quasi-achromatic jet breaks,
while many do not, by associating the former with jets that have
$\gamma_j\theta_j>1$ and the latter with those that have
$\gamma_j\theta_j<1$. Such a scatter in $\gamma_j\theta_j$ might be
naturally produced by differences in GRB environment (affecting
$\theta_j/\theta_{j,c}$, see eq.~\ref{eq_gammathetajnum}) or the
properties of the central engine (affecting $\sigma$). Indeed,
according to \myeqref{eq_gammathetajnum}, a low value of $\sigma$
($\ll1$) or $\theta_j$ ($\ll\theta_{j,c}$) in our jets would mean
$\gamma_j\theta_j<1$ and so the absence of a jet break.

One could use equation~(\ref{eq_gamma_via_mu_sin}),
which is based upon our analytical model of the simulations,
to obtain the Lorentz factor of any GRB jet.
If the black hole or neutron star is nearly maximally spinning with $\Omega\sim 0.25$
and has a polar region with the reasonable value of $\mu\sim 1000$~\citepalias{tchekhovskoy_ff_jets_2008},
and if we consider an opening angle $\theta_j\sim 4^\circ$, then by $r\sim 10^{8}\sim 10^{14}$~cm
one obtains $\gamma_j\sim 250$ and $\gamma_j\theta_j\sim 17$.
For a range of opening angles
with sufficient luminosity, one obtains a range of Lorentz factors
consistent with both short and long duration GRB jets~\citep{pir05, meszaros_grbs_2006}.
Further, the product $\gamma_j\theta_j\gtrsim 1$, indicating
an afterglow can exhibit the so-called ``achromatic jet breaks,''
where observations imply $\gamma_j \theta_j \gtrsim 3$ for long-duration
GRB jets~\citep{pir05,meszaros_grbs_2006}.
Our simulations and analytical models have
$\gamma_j\theta_j \sim 5{-}15$, which is proof of principle that magnetically-driven
jets can produce jet breaks.

\section{Conclusions}
\label{sec_conclusions}

We have studied relativistic magnetized winds from rapidly rotating compact objects
endowed with a split-monopole magnetic field geometry.  We used the relativistic MHD
code, HARM, to simulate these outflows.  We have constructed analytical
approximations to our simulations that describe  fairly accurately the Lorentz
factor and the efficiency of magnetic energy to kinetic energy conversion in the outflow.

Our main result is that, contrary to conventional expectations, the
winds from compact objects endowed with monopole magnetic fields have
{\it efficient} conversion of magnetic energy to kinetic energy near
the rotation axes. We identify this polar wind as a jet since it
contains a sufficiently high luminosity within the required
opening angles of several degrees, and it accelerates
to ultrarelativistic Lorentz factors through an efficient conversion of magnetic energy to
kinetic energy ($\gamma\sim\mu$ and $\sigma<1$ at large radii).
We note that \citet{lyub_eichler_2001} have identified a similar polar jet
in unconfined magnetospheres based on its relative degree of collimation to the
rest of the flow and the relativistic Lorentz factor.  However, they
did not concentrate on the acceleration efficiency and the
transition to the matter-dominated flow. One can use
equation~(\ref{eq_gamma_via_mu_sin}) to show that, for example,
order unity solar mass black holes or neutron stars with $\mu\approx\sigma\sim 300$ near the compact object
will readily produce $\gamma_j\sim 150$ at $10^{14}$ cm with $\theta_j\sim 4^\circ$ such that $\gamma_j\theta_j\sim 10$.

We are able to analytically explain how the jet efficiently converts magnetic energy
to kinetic energy by identifying a ``causality surface,'' beyond
which the jet can no longer communicate with the rotation axis that contains
the flux core.  When one region of the jet can no longer communicate to the
flux core, that region ceases to accelerate efficiently.
The communication between the jet body and the rotation axis
allows magnetic flux surfaces and the
Poynting flux associated with them to become less concentrated
in the main body of the jet (at the expense of the bunch-up near the axis),
and it is this process that allows efficient conversion of magnetic to kinetic energy
(\S\ref{sec_analyt_bpnonuni}).
A similar mechanism, called ``magnetic nozzle,'' was first described
in \citet{begelman_asymptotic_1994}.  We clarify
this mechanism by showing that the accumulation
of flux near the rotation axis leads to a stronger decrease in $\sigma$
for polar field lines than for equatorial field lines.  This effect is
a new feature of ideal MHD winds that has not been discussed by \citet{begelman_asymptotic_1994}
or any other authors except the very recent work by \citet{komi_grb_jets_2008}.

Our results demonstrate that ultrarelativistic jet production is a surprisingly
robust process and probably requires less fine-tuning than previously thought.
It is possible for spinning black holes and neutron stars to produce
ultrarelativistic jets even without the presence of an ambient
confining medium to collimate the jet.  Further,
we show that even unconfined (or weakly confined) winds from compact objects
can produce sufficiently energetic jets ($L_j\sim 10^{49}$ erg/s)
to explain many long GRBs and most short GRBs.

We have confirmed the standard result that monopole magnetospheres are
inefficient accelerators in the equatorial region.  We have thus been unable
to solve the $\sigma$-problem for the Crab PWN
under the assumption of an ideal MHD axisymmetric flow from a star endowed
with a split-monopole magnetosphere.  We note that at some distance from
the neutron star an ideal MHD approximation may break down \citep{usov_1994,
lyutikov_unstable_poynting_2001}.
No highly relativistic jet is observed in the Crab and Vela PWNe, despite
our results that suggest there should be such a feature.  There are observations of
non-relativistic jets with $v/c\sim 0.5$ that appear diffuse and borderline stable.
One way to resolve this discrepancy is that in PWNe systems
the axisymmetric ultrarelativistic jets we find are unstable
to non-axisymmetric perturbations and so can be
a prodigious source of high-energy particles and radiation
via dissipation that causes the jet to slow to non-relativistic velocities \citep[\eg,][]{giannios_jets_2009}.
This notion of a visible jet
emerging directly from the pulsar \citep{lyub_eichler_2001} is an alternative model to the
more recent view that the observed jet is caused by
a post-shock polar backflow with $\sigma\gtrsim 0.01$ that is guided by hoop stresses and forced to
converge toward (and rise up along) the rotation axis \citep{kl2004,zab04}.
In either model, one must consider non-axisymmetric instabilities
since, in the backflow model, the shock structure and backflow could be
highly non-axisymmetric and potentially unstable to non-axisymmetric instabilities.

We remark that while preparing this paper for publication, \citet{komi_grb_jets_2008}
posted a paper describing ideal MHD simulations of confined and unconfined winds.
They do make a minor note that their simulations
of unconfined monopole outflows show efficient conversion of magnetic
energy to kinetic energy near the rotation axis.  We find similar
results to theirs for the unconfined monopole wind.  We are further
able to make analytical estimates that explain the nature of this
efficient conversion via introducing a causality surface at which
the jet loses causal connection with the polar axis.
We are also able to obtain a closed-form approximation for the
Lorentz factor in this region based upon a precise notion
of the ``causality surface.''  We note that previous authors who studied
stars endowed with a split-monopole field geometry in the ideal MHD
approximation \citep{bog01,buc06}
did not perform simulations to large enough radii in order to observe
the outflow achieving such an efficient conversion of magnetic to kinetic energy
near the rotation axis or reaching such large Lorentz factors.

We conclude with prospects for future research.
Our solutions are axisymmetric,
assume the compact object is endowed with a monopolar field geometry,
assume particles are injected with $\gamma\sim 1$ near the compact object,
assume $\sigma\gg 1$ near the equatorial plane,
and assume the ideal cold MHD approximation holds.
In future work we plan to consider the stability of our solutions to non-axisymmetric instabilities
\citep[\cf][]{nlt09},
the effect of external confinement leading to highly-collimated solutions,
the injection of ultrarelativistic particles as suggested to occur, e.g., in the Crab pulsar,
the injection of non-relativistic particles as may be relevant for thin disks
\citep{bog99},
and the effect of $\sigma\sim 0$ near the equatorial plane
as required near the pulsar's equatorial current sheet or
for any system with an accretion disk.

\acknowledgements

We thank Omer Blaes and Anatoly Spitkovsky for useful discussions,
Niccolo' Bucciantini for helpful comments
and the anonymous referee for a detailed review
that helped to improve the manuscript.
The simulations described in this paper were run on the Odyssey
cluster supported by the Harvard FAS Research Computing Group and the
BlueGene/L system at the Harvard SEAS CyberInfrastructures Lab.
This work was supported in part by NASA grant NNX08AH32G, by the
National Science Foundation through TeraGrid resources~\citep{catlett2007tao} provided by the
Louisiana Optical Network Initiative (\href{http://www.loni.org}{http://www.loni.org}), and
by NASA's Chandra Fellowship PF7-80048 (JCM).

\appendix

\section{Lorentz factor in an ideal MHD flow}
\label{sec_gamma_mhd}

In this section we derive an approximate expression for the Lorentz
factor in an ideal MHD flow.
Conservation of energy and angular momentum flux along a field line
imply~\citep[\eg,][]{con02}:
\begin{equation}
\gamma(1-\Omega R v_\varphi) = \gamma_0,
\label{eq_energy_angular}
\end{equation}
where $\Omega$ is the angular frequency of field line and
$\gamma_0$ and $\gamma$ are the initial and the local
Lorentz factors of the field line,
\begin{equation}
\gamma = (1-v_p^2-v_\varphi^2)^{-1/2}.
\label{eq_gamma}
\end{equation}
For convenience, introduce auxiliary variables $x=\Omega R$ and $y$:
\begin{eqnarray}
E &=& x B_p, \\
B_\varphi &=& -y B_p. \label{eq_y}
\end{eqnarray}
The drift Lorentz factor~\myeqref{eq_gammadrift} is
\begin{equation}
\gamma_\mytext{dr}^2 = \frac{B^2}{B^2-E^2} = \frac{1+y^2}{1+y^2-x^2}.
\label{eq_gammadrift_appendix}
\end{equation}
Resolving \myeqref{eq_bc_vphi}, \myeqref{eq_energy_angular}--\myeqref{eq_y} for
$\gamma$, we obtain:
\begin{equation}
\gamma^2=\frac{\left[\gamma _0^2\left(1+y^2\right)+x^2y^2\right]^2}
     {\left(1+y^2-x^2\right)
        \left[xy(x^2+\gamma _0^2-1)^{1/2}+\gamma_0(1+y^2-x^2)^{1/2}\right]^2}.
        \label{eq_gamma_exact_mhd}
\end{equation}
This equation is exact and expresses the MHD fluid velocity via the
local parameters ($x$, $y$) and the footpoint parameters ($\gamma_0$).
In the limit $x \gg 1$, we have the following
analytic approximation
for the MHD Lorentz factor  (assuming that $\abs{B_\varphi}\approx E$, or that
$y^2-x^2\ll x^2$),
\begin{equation}
\gamma_a^2 = \frac{\gamma_0^2+x^2}{1+y^2-x^2},
\label{eq_gammaa}
\end{equation}
that is very close to the drift Lorentz factor~\myeqref{eq_gammadrift_appendix}.
This shows that asymptotically the Lorentz factor of the MHD
flow~\myeqref{eq_gamma_exact_mhd} is closely approximated by
the drift Lorentz factor~\myeqref{eq_gammadrift_appendix}
\citep[\cf][]{bes98,beszak04,vla04}. We note that
another convenient, albeit slightly less accurate,
form of~\myeqref{eq_gammaa} is
\begin{equation}
\gamma_a^2 \approx \gamma_0^2-1+\gamma_\mytext{dr}^2.
\end{equation}

As we show by comparison to numerical results,
formula~\myeqref{eq_gammaa} works extremely well not only asymptotically
but at all distances from the star.  We now show analytically
why this is the case.  In the monopole flow
$\abs{B_\varphi} \approx E$ at all distances from
the star (see \S\ref{sec_analyt_aoafp}), therefore neglecting
$y^2-x^2$ as compared to $x^2$, we obtain:
\begin{equation}
\frac{\gamma}{\gamma_a}
   \approx \frac{\gamma_0^2(1+x^2)+x^4}%
   {(\gamma_0^2+x^2)^{1/2}\left[\gamma_0 +x^2(\gamma_0^2+x^2-1)^{1/2}\right]}.
\end{equation}
According to this equation, the relative deviation of $\gamma_a$
\wrt\ $\gamma$ is always smaller than $6$\% for any
$x\equiv \Omega R$ and $\gamma_0$.
Therefore, formula~\myeqref{eq_gammaa} can be used as an accurate approximation
for the MHD fluid Lorentz factor at all distances from the star and
initial Lorentz factor $\gamma_0$.

We now recast the approximation to the Lorentz factor
$\gamma_a$~\myeqref{eq_gammaa} in a two-component form \citepalias[\cf][]{tchekhovskoy_ff_jets_2008}:
\begin{equation}
\frac{1}{\gamma^2} \approx \frac{1}{\gamma_1^2} +\frac{1}{\gamma_2^2},
                   \label{eq_gamma_two_components}
\end{equation}
where
\begin{equation}
\gamma_1^2 = \gamma_0^2 + x^2 = \gamma_0^2 + (\Omega R)^2
\end{equation}
increases roughly linearly with distance and
\begin{equation}
\gamma_2^2 \approx \frac{x^2}{y^2-x^2} = \frac{E^2}{B_\varphi^2-E^2}
\end{equation}
is related to the poloidal radius of curvature of the
field line~$R_c$ \citepalias{tchekhovskoy_ff_jets_2008} in the limit $\gamma_2 \ll \mu$:
\begin{equation}
\gamma_2^2 \approx C  \frac{R_c}{R} \cos\theta
\label{eq_RcoR}
\end{equation}
where $C\simeq1$ is a numerical factor.
Since $\gamma^2/2$ is the harmonic mean of $\gamma_1^2$ and
$\gamma_2^2$, the \emph{smaller} of $\gamma_1$ and $\gamma_2$
determines the total Lorentz factor. Close to the compact object the first
term in~\myeqref{eq_gamma_two_components} dominates, and $\gamma$ increases
roughly linearly with distance from the axis (see \S\ref{sec_causality} and
\citetalias{tchekhovskoy_ff_jets_2008}),
\begin{equation}
\gamma \approx \gamma_1 = \left[\gamma_0^2 + (\Omega R)^2\right]^{1/2} \approx \Omega R.
\end{equation}
As we discuss in \S\ref{sec_causality}, beyond the causality surface
the second term in~\myeqref{eq_gamma_two_components}
becomes dominant and there
the Lorentz factor grows logarithmically,
\begin{equation}
\gamma \approx \gamma_2 \propto \ln^{1/3}r.
\label{eq_logthird_appendix}
\end{equation}
In the next section we present a compact derivation of this result.

\section{Logarithmic acceleration regime}
\label{sec_logthird}

In this section we consider acceleration along near-midplane
field lines, which have $\theta' = \pi/2 - \theta \ll 1$,
and show that it is logarithmic sufficiently
far from the central star where the flow has switched
to the second acceleration regime~\myeqref{eq_RcoR}.
We limit ourselves to a highly magnetized region of the flow, $\gamma\ll\mu$.
Assuming a near-monopolar structure
in this region, we express the rate of change of angle $\theta'$
along a field line in terms of its local poloidal curvature radius $R_c$ as
\begin{equation}
\frac{d\theta'}{dr} \approx \frac{1}{R_c}
                    {\approx}
                    \frac{C \theta'}{r \gamma^2},
\label{eq_dthdr}
\end{equation}
where in the latter equality we used equation~\myeqref{eq_RcoR}
and the fact that $r\approx R$ for $\theta'\ll1$.
Differentiating equation~\myeqref{eq_eff_near_midplane}, we express $d\gamma/d\theta'$
along the field line:
\begin{equation}
\frac{d\gamma}{d\theta'} \approx \mu \frac{\theta'_\fpt}{(\theta')^2}.
\label{eq_dgdth}
\end{equation}
Combining equations~\myeqref{eq_dthdr} and \myeqref{eq_dgdth}
and assuming a near-monopolar structure, $\theta' \approx \theta'_\fpt$,
we obtain a differential equation for $\gamma$:
\begin{equation}
\frac{\gamma^2d\gamma}{\mu} \approx C \frac{dr}{r},
\end{equation}
or
\begin{equation}
\gamma \approx (C \mu)^{1/3} \ln^{1/3}(r/r_*),
\end{equation}
where $r_*$ is an integration constant.  In~\S\ref{sec_analyt_approx}
we argue that $r_* \approx r_c$, the radius at which the field line
intersects the causality surface.

\section{Auxiliary approximate analytic solution}
\label{sec_aux_approx_sol}

HARM integrates the ideal cold MHD equations of motion by first interpolating
the primitive quantities (i.e. density, velocity, and magnetic field) from cell centers to cell faces.  The
cell faces will generally now have two values corresponding to an interpolation
from different cell centers.  As described in \citet{gam03}, these two values
are used to compute a generally dissipative flux that is used to advance
the set of conserved quantities forward in time.  These conserved quantities
are then inverted to produce the new primitive quantities.

In cases where one roughly knows the solution for the primitive quantities
as functions of position, one can instead interpolate the ratio of the primitive
quantities to the estimated solution.  This reduces numerical dissipation by reducing
the difference in the two values of primitive quantities interpolated to the face.
The closer the estimated solution is to the true solution, the lower the numerical
dissipation.  So it is useful to estimate the solution as best one can.  The particular
form of this estimated solution has no other effect on the numerical simulation.

Here we provide our estimated solution based upon an approximate extension of
known force-free solutions \citepalias{tchekhovskoy_ff_jets_2008}:
\begin{eqnarray}
B_r &=& r^{-2},\\
B_\theta &=& 0, \\
B_\varphi &=&  - \Omega R B_r, \\
\frac{1}{\gamma^2-\gamma_0^2} &=& \frac{1}{(\Omega R)^2}+\frac{1}{(\mu/2)^2},\label{gammafit} \\
u_r^2 &=& \gamma^2 - 1, \\
u_\theta &=& 0, \\
u_\varphi &=& \frac{\gamma \Omega R B_r^2}{B_r^2+B_\varphi^2} ,\\
\rho &=& \eta \frac{B_r}{u_r},
\end{eqnarray}
with vectors given in an orthonormal (physical) basis and
where $\eta$ is given by the boundary conditions according to equation~(\ref{eq_etadef}).
We interpolate the numerical solution (given in a contravariant coordinate basis)
divided by the estimated solution (also in a contravariant coordinate basis),
with the exception of $B_\theta$ and $u_\theta$ for which we do not divide
by the analytic solution.
In addition, we independently interpolate $\gamma$ using equation~(\ref{gammafit})
in order to rescale the interpolated 4-velocity,
which leads to a more accurate solution \citep{tch_wham07}.

\section{Mach Cone for Fast Magnetosonic Waves}
\label{sec_mach_cone}

In this section we derive the expression for the half-opening angle
of a Mach cone for fast magnetosonic waves.
Consider a segment of the relativistic magnetized wind propagating
with a velocity vector $\myvec{\mybeta}$ and Lorentz factor $\gamma$.
Let a fast
magnetosonic wave travel in the comoving fluid frame at an angle
$\xi'$ to the direction $\myvec{\mybeta}$. Decomposing the wave velocity
into parallel and perpendicular components, we have
\begin{equation}
\mybeta'_{\parallel}=\mybeta_f\cos\xi', \qquad
\mybeta{'}_\perp=\mybeta_f\sin\xi'.
\end{equation}
Transforming back to the lab frame, the components of the wave
velocity become
\begin{equation}
\mybeta_{\parallel}=\frac{\mybeta_f\cos\xi'+\mybeta}{(1+\mybeta\mybeta_f\cos\xi')}, \qquad
\mybeta_\perp=\frac{\mybeta_f\sin\xi'}{\gamma(1+\mybeta\mybeta_f\cos\xi')}.
\end{equation}
Therefore, in the lab frame, the wave velocity vector is oriented at
an angle $\xi$ to the vector $\myvec{\mybeta}$, where
\begin{equation}
\tan\xi = \frac{\mybeta_{\perp}}{\mybeta_{\parallel}} =
\frac{\mybeta_f\sin\xi'}{\gamma(\mybeta_f\cos\xi'+\mybeta)}.
\end{equation}

In the lab frame, different waves have different values of $\xi$.
Assuming that the medium moves with a superfast speed (i.e.,
$\mybeta>\mybeta_f$), all waves move in the downstream direction in the
lab frame and their $\xi$ values are all less than a certain maximum
$\xi_{\rm max}$.  To determine $\xi_{\rm max}$, we maximize $\tan\xi$
with respect to variations of $\xi'$.  The maximum is achieved when
\begin{equation}
\cos\xi'=-\frac{\mybeta_f}{\mybeta}, \qquad
\sin\xi' = \frac{\sqrt{\mybeta^2-\mybeta_f^2}}{\mybeta}.
\end{equation}
This gives the half-opening angle of the Mach cone, $\xi_{\rm max}$:
\begin{eqnarray}
\tan\xi_{\rm max} &=& \frac{\mybeta_f}{\gamma\sqrt{\mybeta^2-\mybeta_f^2}}, \\
\label{tanximax}
\sin\xi_{\rm max} &=& \frac{\gamma_f \mybeta_f}{\gamma \mybeta} = \frac{1}{M_f}.
\end{eqnarray}


\label{lastpage}
\end{document}